\documentclass[prd,twocolumn,amsmath,amssymb,floatfix,nofootinbib]{revtex4}
\usepackage{mathrsfs,bm}
\usepackage{longtable,lscape}
\usepackage{txfonts}
\usepackage{indentfirst}
\usepackage{graphicx,color,dcolumn,booktabs}
\usepackage{multirow}
\usepackage{indentfirst}
\usepackage{multirow}
\usepackage{epsfig}
\usepackage{graphicx,color,dcolumn}
\usepackage{epstopdf}
\usepackage{amsmath}
\usepackage{multirow}
\usepackage{diagbox}
\usepackage{changes}
\usepackage{threeparttable}
\usepackage{enumitem}

\usepackage{color}
\usepackage{amssymb}
\definecolor{cover}{rgb}{0.77,0.87,0.88}
\definecolor{blueone}{rgb}{0.1,0.1,.7}
\definecolor{citec}{rgb}{0.14,0.47,0.09}
\definecolor{two}{rgb}{0.0,0.5,0.}
\definecolor{three}{rgb}{.5,.1,0.15}
\usepackage[bookmarks=true,bookmarksopen=false,plainpages=false,breaklinks=true,
   bookmarksnumbered=true,hypertexnames=false,
   filecolor=blue,urlcolor=three,menucolor=three,
   linkcolor=three,citecolor=blueone, colorlinks,
   anchorcolor=blue,runcolor=pink,frenchlinks=red
   pdfstartview=FitH,pdftitle=title,%
   pdfauthor=author]{hyperref}

\usepackage{relsize}
\usepackage{xspace}
\def\babar{\mbox{\slshape B\kern-0.1em{\smaller A}\kern-0.1em
    B\kern-0.1em{\smaller A\kern-0.2em R}}}
\newcommand{\myvec}[1]{\ensuremath{\begin{pmatrix}#1\end{pmatrix}}}

\allowdisplaybreaks[4]
\usepackage{color}
\begin{document}
\title{Systematical study of $\Omega_c$-like molecular states   from interactions  $\Xi_c^{(',*)}\bar{K}^{(*)}$ and $\Xi^{(*)}D^{(*)}$}

\author{Jun-Tao Zhu, Shu-Yi Kong, Lin-Qing Song, and Jun He\footnote{Corresponding author: junhe@njnu.edu.cn}}
\affiliation{School of Physics and Technology, Nanjing Normal University, Nanjing 210097, China}

\date{\today} 

\begin{abstract} In this work, the  $\Omega_c$-like molecular states are
systematically investigated in a quasipotential Bethe-Salpeter equation
approach. The relevant interactions $\Xi_c^{(*,')}\bar{K}^{(*)}$,
$\Xi^{(*)}D^{(*)}$, and $\Omega^{(*)}_c(\pi/\eta/\rho/\omega)$ are described by
light meson exchanges with the help of the effective Lagrangians with SU(3),
chiral, and heavy quark symmetries.  The obtained potential kernels of
considered interactions are inserted into the quasipotential Bethe-Salpeter
equation, and coupled-channel calculations are performed to find possible
molecular states and its couplings to the channels considered. The results
suggest that an isoscalar state can be produced from the $\Xi^*_c\bar{K}$
interaction  with spin parity $3/2^-$, which can be related to state
$\Omega_c(3120)$. And its isoscalar partner is predicted with a dominant decay
in the $\Omega_c^*\pi$ channel. The isoscalar and isovector states  with $1/2^-$
can be produced from the $\Xi'_c\bar{K}$ interaction with a threshold close to
the mass of the $\Omega_c(3050)$ and $\Omega_c(3065)$.  Their couplings to the
$\Xi_c \bar{K}$ channel  are very weak, and the isovector one has strong
coupling to $\Omega_c\pi$.  High-precision measurement is helpful to confirm or
search such molecular states. Experimental search of states with higher masses
generated from interactions $\Xi^{(*,')}_c \bar{K}^*$ and $\Xi^* D^{(*)}$ are
also suggested by the current results.


\end{abstract}

\maketitle
\section{INTRODUCTION}\label{sec1}

In 2017, LHCb reported five  narrow structures named $\Omega_c (3000)$,
$\Omega_c (3050)$, $\Omega_c (3065)$, $\Omega_c (3090)$, and $\Omega_c (3120)$
in the $\Xi_c^+ K^-$ mass projection of the $\Omega_b^-\rightarrow\Xi_c^+
K^-\pi^-$ decays~\cite{LHCb:2017uwr}. The Belle Collaboration confirmed the former
four structures~\cite{Belle:2017ext}.   Many theoretical works were inspired by
the observations to interpret their origins and internal structures, including
calculations in the potential
model~\cite{Chiladze:1997ev,Chen:2017xat,Nieves:2017jjx,Montana:2017kjw,Debastiani:2017ewu},
constituent quark
model~\cite{Wang:2017vnc,Chen:2017gnu,Luo:2021dvj,Galkin:2020iat,Roberts:2007ni,Shah:2016mig,Yoshida:2015tia,Karliner:2017kfm,Wang:2017hej,Agaev:2017lip,Santopinto:2018ljf,Kim:2017jpx,An:2017lwg,Huang:2017dwn},
QCD sum rule~\cite{Wang:2017zjw,Chen:2017sci,Chen:2015kpa}, lattice
QCD~\cite{Padmanath:2017lng}, and other phenomenological
approaches~\cite{Cheng:2017ove,Zhao:2017fov}.  One of the most popular
interpretations is that the observed five peaks correspond to five excited
$\Omega_c$ baryons with spin parities $J^P=1/2^-$, $1/2^-$, $3/2^-$, $3/2^-$,
and $5/2^-$, respectively. The $\Omega_c(3120)$ and
$\Omega_c(3050)/\Omega_c(3065)$ are close to the $\Xi^*_c\bar{K}$ and
$\Xi'_c\bar{K}$ thresholds, respectively.  The molecular states picture was also
applied to interpret some of five peaks as baryon-meson bound
states~\cite{Chen:2017xat,Nieves:2017jjx, Kim:2017jpx, An:2017lwg, Montana:2017kjw,
Huang:2017dwn,Debastiani:2017ewu}.  For example, in the chiral unitary approach,
the theoretical masses and widths of a $\Xi^*_c\bar{K}$ state with $3/2^-$,
a $\Xi D$ state with $1/2^-$, and a $\Xi'_c\bar{K}$ state with $1/2^-$ are in
remarkable agreement with the experimentally observed $\Omega_c (3120)$,
$\Omega_c (3090)$, and $\Omega_c (3050)$~\cite{Debastiani:2017ewu}.  In
Ref.~\cite{Chen:2017xat}, a coupled-channel calculation indicates that either
$\Omega_c (3090)$ or $\Omega_c (3120)$ is possibly  to be related to isoscalar
$\Xi^*_c\bar{K}/\Omega_c\eta/\Omega_c^*\eta/\Xi_c\bar{K}^*/\Xi'_c\bar{K}^*/\Omega_c\omega$
state with spin parity $3/2^-$.

In 2021, the LHCb Collaboration updated its measurement about the $\Omega_c$
structures~\cite{LHCb:2021ptx}. The new analysis suggests that the spins of
$\Omega_c(3050)$ and $\Omega_c(3065) $ tend not to be  $1/2$. However,
$\Omega_c(3050)$ and $\Omega_c(3065) $ are close to the $\Xi'_c\bar{K}$ threshold.
If we assign them as molecular states composed
of the $\Xi'_c\bar{K}$, only spin parity $1/2^-$ can be obtained in the $S$ wave.  Hence,
the new result is inconsistent with the prediction of spin of the
$\Omega_c(3050) $ in previous theoretical
works~\cite{Karliner:2017kfm,Wang:2017zjw,Padmanath:2017lng,Debastiani:2017ewu}.
In the Belle experiment~\cite{Belle:2017ext} and new experiment at
LHCb~\cite{LHCb:2021ptx}, the $\Omega_c(3120)$ was not observed, though it is a
good candidate of $\Xi^*_c\bar{K}$ molecular state with $3/2^-$. In
Ref.~\cite{LHCb:2021ptx}, it was suggested that   the $\Omega_c(3120)$
would be a state being either one of the $2S$ doublet, or a $\rho$-mode $P$-wave
excitation, which decays to $\Xi_c^+ K^-$ in the $D$ wave, and then suppressed.

Up to now, the internal structures of these $\Omega_c$-like states are still not
well understood. We should answer why the $\Omega_c(3120)$ is so difficult to
observe if it is a molecular state. If we accept the new LHCb results about the
spin parities, the $\Xi'_c\bar{K}$ molecular states should not be 
$\Omega_c(3050)$ or $\Omega_c(3065)$. Where should we find them?  In this work,
we will investigate all $\Omega_c$-like baryon-meson interactions with flavor
numbers $C=1$, $S=-2$ under $3.6$~GeV to find all possible molecular states from
these interactions and discuss their relations to the experimentally observed
structures. The couplings of these molecular states to the channels considered
will also be studied through coupled-channel calculation. Besides, more
molecular states will be  predicted from these interactions, which are helpful
to understand existing states and future experimental research.

Based on such consideration, we will consider interactions
$\Xi_c^{(*,')}\bar{K}^{(*)}$, $\Xi^{(*)}D^{(*)}$, and
$\Omega^{(*)}_c(\eta/\omega/\pi/\rho)$ in the current work. As usual, only
spin parities $J^P$ of the interactions which can be produced in the $S$-wave will be
included in the calculation. However, we would like to note that
contributions from higher partial waves will also be included in our models for
an interaction considered.  All possible isospins and spin parities of all
considered interactions are listed in Table~\ref{T1} in order of mass threshold
from large to small. Eighteen interactions are considered in the current work,
and 48 channels will be involved after different isospins and spin parities are
considered.  In the calculation, these interactions can be divided into two categories:
\begin{itemize}[itemsep= 0 pt]
 \item category I: $\Xi^{(*)}D^{(*)}$;
 \item category II: $\Xi_c^{(*,')}\bar{K}^{(*)}$,~$\Omega^{(*)}_c(\pi/\eta/\rho/\omega)$.
\end{itemize}
In the current work, the one-boson-exchange model will be adopted to describe
the interactions. The interactions in category I are composed of a light baryon
and a charmed meson, while the interactions in category II are composed of a
charmed baryon and a light meson. The couplings between channels belonging to
different categories intermediate only by exchanges of charmed mesons.  It
should be heavily suppressed and can be ignored compared with the couplings
between the channels in the same category where  light meson exchanges
provide the dominant contribution.  Besides, the
$\Omega^{(*)}_c(\pi/\eta/\rho/\omega)$ interaction in category II will be
excluded in single-channel calculation  in the current model due to the absence
of possible meson exchange to provide attraction.  However, these channels are
not trivial in coupled-channel calculation.

\renewcommand\tabcolsep{0.075cm}
\renewcommand{\arraystretch}{1.5}
 \begin{table}[h!] \footnotesize \caption{ The
possible isospins and spin parities  of all considered interactions. The
thresholds are in the unit of MeV.}\label{T1} \centering
\begin{tabular}{c|cccccc}\toprule[1pt] {Channel}&$\Omega^*_c\omega$& $ \Xi
^*D^*$&$\Omega_c^*\rho$&$\Xi^*_c\bar{K}^*$&$\Omega_c\omega$&$\Xi'_c\bar{K}^*$\\
{ Threshold}&   $3550.9$       &$3541.8$    & $3541.2$                 &$3540.2$
&$3480.1$ &$3470.7$\\ 
$I=0,J^P$&$(\frac{1}{2},\frac{3}{2},\frac{5}{2})^-$
&$(\frac{1}{2},\frac{ 3}{2},\frac{5}{2})^-$  & $--$
&$(\frac{1}{2},\frac{ 3}{2},\frac{5}{2})^-$  & $(\frac{1}{2},\frac{ 3}{2})^-$
&$(\frac{1}{2},\frac{ 3}{2})^-$ \\
{$I=1,J^P$}&$--$ &$(\frac{1}{2},\frac{ 3}{2},\frac{5}{2})^-$
& $(\frac{1}{2},\frac{ 3}{2},\frac{5}{2})^-$
& $(\frac{1}{2},\frac{ 3}{2},\frac{5}{2})^-$
&$--$&$(\frac{1}{2},\frac{ 3}{2})^-$\\ \hline

{Channel}&$\Omega_c\rho$ &$\Xi ^*D$ &$\Xi_c \bar{K}^*$&  $\Xi D^*$& $\Omega^*_c \eta$&$\Omega_c \eta$\\
{ Threshold}&$3470.5$& $3400.6$  & $3363.3$ & $3326.5$& $3315.8$& $3245.0$\\
{$I=0,J^P$} & $--$& $\frac{ 3}{2}^-$&$(\frac{1}{2},\frac{ 3}{2})^-$ &$(\frac{1}{2},\frac{ 3}{2})^-$ &$\frac{ 3}{2}^-$& $\frac{1}{2}^-$\\ 
{$I=1,J^P$} &$(\frac{1}{2},\frac{ 3}{2})^-$& $\frac{ 3}{2}^-$ & $(\frac{1}{2},\frac{ 3}{2})^-$&$(\frac{1}{2},\frac{ 3}{2})^-$  &$--$& $--$ \\ \hline

{Channel} &$\Xi D$                 & $\Xi^*_c\bar{K}$     & $\Xi'_c\bar{K}$  &$\Xi_c\bar{K}$         &$\Omega_c^*\pi$ &$\Omega_c\pi$\\
{Threshold} & $3185.3$      & $3142.0$                      &$3072.5$               &$2965.1$                  &$2903.1$                   &$2832.4$\\ 
{$I=0,J^P$} &$\frac{1}{2}^-$ & $\frac{3}{2}^-$ & $\frac{1}{2}^-$&$\frac{1}{2}^-$            &$--$                             &$--$\\ 
{$I=1,J^P$}&$\frac{1}{2}^-$ & $\frac{3}{2}^-$ & $\frac{1}{2}^-$&$\frac{1}{2}^-$              &$\frac{3}{2}^-$      &$\frac{1}{2}^-$\\
\bottomrule[1pt]
\end{tabular}
\end{table}

This article is organized as follows.  In the next section, the effective
Lagrangians will be provided to construct  potentials of considered
interactions $\Xi^{(*)}D^{(*)}$, $\Xi_c^{(*,')}K^{(*)}$, and
$\Omega^{(*)}_c(\pi/\eta/\rho/\omega)$ in the one-boson-exchange model. The
quasipotential Bethe-Salpeter equation approach adopted in the current work will
be also introduced briefly in that section.  The results with single-channel
calculation will be presented in  section~\ref{sec3}. And  couplings between
different interactions will be specifically described and discussed in section~\ref{sec4}.
Finally, the article ends with summary and discussion in section
~\ref{sec5}.

\section{Theoretical frame}\label{sec2}

All systems considered in the current work are composed of a charmed and a light
hadron. The Lagrangians for the charmed and light hadrons will be presented in the following.

\subsection{Lagrangians for charmed hadrons}

The Lagrangians for the vertex of charmed hadrons and light mesons have been
constructed under the heavy quark limit and chiral symmetry in the literature
\cite{Cheng:1992xi,Yan:1992gz,Wise:1992hn,Casalbuoni:1996pg}.  The couplings of
heavy-light charmed mesons $\mathcal{P}^{(*)}=(D^{(*)0}, D^{(*)+},D^{(*)+}_s)$  and light exchange
mesons can be depicted as
\begin{align}
  \mathcal{L}_{\mathcal{P}^*\mathcal{P}\mathbb{P}} &=
 i\frac{2g\sqrt{m_{\mathcal{P}} m_{\mathcal{P}^*}}}{f_\pi}
  (-\mathcal{P}^{*\dag}_{a\lambda}\mathcal{P}_b
  +\mathcal{P}^\dag_{a}\mathcal{P}^*_{b\lambda})
  \partial^\lambda\mathbb{P}_{ab},\nonumber\\
    \mathcal{L}_{\mathcal{P}^*\mathcal{P}^*\mathbb{P}} &=
-\frac{g}{f_\pi} \epsilon_{\alpha\mu\nu\lambda}\mathcal{P}^{*\mu\dag}_a
\overleftrightarrow{\partial}^\alpha \mathcal{P}^{*\lambda}_{b}\partial^\nu\mathbb{P}_{ba},\nonumber\\
    \mathcal{L}_{\mathcal{P}^*\mathcal{P}\mathbb{V}} &=
\sqrt{2}\lambda g_V\varepsilon_{\lambda\alpha\beta\mu}
  (-\mathcal{P}^{*\mu\dag}_a\overleftrightarrow{\partial}^\lambda
  \mathcal{P}_b  +\mathcal{P}^\dag_a\overleftrightarrow{\partial}^\lambda
 \mathcal{P}_b^{*\mu})(\partial^{\alpha}\mathbb{V}^\beta)_{ab},\nonumber\\
	\mathcal{L}_{\mathcal{P}\mathcal{P}\mathbb{V}} &= -i\frac{\beta	g_V}{\sqrt{2}}\mathcal{P}_a^\dag
	\overleftrightarrow{\partial}_\mu \mathcal{P}_b\mathbb{V}^\mu_{ab}, \nonumber\\
  \mathcal{L}_{\mathcal{P}^*\mathcal{P}^*\mathbb{V}} &=  i\frac{\beta
  g_V}{\sqrt{2}}\mathcal{P}_a^{*\dag}\overleftrightarrow{\partial}_\mu
  \mathcal{P}^*_b\mathbb{V}^\mu_{ab}\nonumber\\
  &-i2\sqrt{2}\lambda  g_Vm_{\mathcal{P}^*}\mathcal{P}^{*\mu\dag}_a\mathcal{P}^{*\nu}_b(\partial_\mu\mathbb{V}_\nu-\partial_\nu\mathbb{V}_\mu)_{ab}
,\nonumber\\
  \mathcal{L}_{\mathcal{P}\mathcal{P}\sigma} &=
  -2g_s m_{\mathcal{P}}\mathcal{P}_a^\dag \mathcal{P}_a\sigma, \nonumber\\
  \mathcal{L}_{\mathcal{P}^*\mathcal{P}^*\sigma} &=
  2g_s m_{\mathcal{P}^*}\mathcal{P}_a^{*\dag}
  \mathcal{P}^*_a\sigma,\label{LD}
\end{align}
where $m_{\mathcal{P}^{(*)}}$ is the mass of  $\mathcal{P}^{(*)}$ and $\overleftrightarrow{\partial}=\overrightarrow{\partial}-\overleftarrow{\partial}$. The
$\mathcal{P}
$ and $\mathcal{P}^*
$ satisfy the normalization relations $\langle
0|{\mathcal{P}}|\bar{Q}{q}(0^-)\rangle
=\sqrt{m_\mathcal{P}}$ and $\langle
0|{\mathcal{P}}^*_\mu|\bar{Q}{q}(1^-)\rangle=
\epsilon_\mu\sqrt{m_{\mathcal{P}^*}}$.
The $\mathbb
P$ and $\mathbb V$ are the pseudoscalar and vector matrices, 
\begin{align}
 {\mathbb P}=\left(\begin{array}{ccc}
      \frac{\sqrt{3}\pi^0+\eta}{\sqrt{6}}&\pi^+&K^+\\
      \pi^-&\frac{-\sqrt{3}\pi^0+\eta}{\sqrt{6}}&K^0\\
      K^-&\bar{K}^0&\frac{-2\eta}{\sqrt{6}}
\end{array}\right),
\mathbb{V}=\left(\begin{array}{ccc}
\frac{\rho^0+\omega}{\sqrt{2}}&\rho^{+}&K^{*+}\\
\rho^{-}&\frac{-\rho^{0}+\omega}{\sqrt{2}}&K^{*0}\\
K^{*-}&\bar{K}^{*0}&\phi
\end{array}\right).\label{MPV}
\end{align}

The Lagrangians for the couplings between  charmed baryon and light exchanged
meson can also be constructed under the heavy quark limit and  chiral symmetry, and the explicit forms of these Lagrangians can be written as~\cite{Liu:2011xc}
\begin{align}
{\cal L}_{BB\mathbb{P}}&=-\frac{3g_1}{4f_\pi\sqrt{m_{\bar{B}}m_{B}}}~\epsilon^{\mu\nu\lambda\kappa}\partial^\nu \mathbb{P}~
\sum_{i=0,1}\bar{B}_{i\mu} \overleftrightarrow{\partial}_\kappa B_{j\lambda},\nonumber\\
{\cal L}_{BB\mathbb{V}}&=-i\frac{\beta_S g_V}{2\sqrt{2m_{\bar{B}}m_{B}}}\mathbb{V}^\nu
 \sum_{i=0,1}\bar{B}_i^\mu \overleftrightarrow{\partial}_\nu B_{j\mu}\nonumber\\
&-i\frac{\lambda_Sg_V}{\sqrt{2}}(\partial_\mu \mathbb{V}_\nu-\partial_\nu \mathbb{V}_\mu) \sum_{i=0,1}\bar{B}_i^\mu B_j^\nu,\nonumber\\
{\cal L}_{BB\sigma}&=\ell_S\sigma\sum_{i=0,1}\bar{B}_i^\mu B_{j\mu},\nonumber\\
    {\cal L}_{B_{\bar{3}}B_{\bar{3}}\mathbb{V}}&=-i\frac{g_V\beta_B}{2\sqrt{2m_{\bar{B}_{\bar{3}}}m_{B_{\bar{3}}}} }\mathbb{V}^\mu\bar{B}_{\bar{3}}\overleftrightarrow{\partial}_\mu B_{\bar{3}},\nonumber\\
{\cal L}_{B_{\bar{3}}B_{\bar{3}}\sigma}&=\ell_B \sigma \bar{B}_{\bar{3}}B_{\bar{3}},\nonumber\\
{\cal L}_{BB_{\bar{3}}\mathbb{P}}&=-i\frac{g_4}{f_\pi} \sum_i\bar{B}_i^\mu \partial_\mu \mathbb{P} B_{\bar{3}}+{\rm H.c.},\nonumber\\
{\cal L}_{BB_{\bar{3}}\mathbb{V}} &=\frac{g_\mathbb{V}\lambda_I}{\sqrt{2m_{\bar{B}}m_{B_{\bar{3}}}}} \epsilon^{\mu\nu\lambda\kappa} \partial_\lambda \mathbb{V}_\kappa\sum_i\bar{B}_{i\nu} \overleftrightarrow{\partial}_\mu
   B_{\bar{3}}+{\rm H.c.}.
   \label{LB}
\end{align}
where $S^{\mu}_{ab}$ read
\begin{align}
{ B}^{ab}_{0\mu}&=-\sqrt{\frac{1}{3}}(\gamma_{\mu}+v_{\mu})
    \gamma^{5}B^{ab}, \ \
B^{ab}_{1\mu}=B^{*ab}_{\mu},\nonumber\\
{\bar{B}}^{ab}_{0\mu}&=\sqrt{\frac{1}{3}}\bar{B}^{ab}
    \gamma^{5}(\gamma_{\mu}+v_{\mu}), \  \ \  \
\bar{B}^{ab}_{1\mu}=\bar{B}^{*ab}_{\mu},
\end{align}
with the charmed baryon matrices being defined as
\begin{align}
B_{\bar{3}}&=\left(\begin{array}{ccc}
0&\Lambda^+_c&\Xi_c^+\\
-\Lambda_c^+&0&\Xi_c^0\\
-\Xi^+_c&-\Xi_c^0&0
\end{array}\right),\quad
B=\left(\begin{array}{ccc}
\Sigma_c^{++}&\frac{1}{\sqrt{2}}\Sigma^+_c&\frac{1}{\sqrt{2}}\Xi'^+_c\\
\frac{1}{\sqrt{2}}\Sigma^+_c&\Sigma_c^0&\frac{1}{\sqrt{2}}\Xi'^0_c\\
\frac{1}{\sqrt{2}}\Xi'^+_c&\frac{1}{\sqrt{2}}\Xi'^0_c&\Omega^0_c
\end{array}\right). \nonumber\\
B^*&=\left(\begin{array}{ccc}
\Sigma_c^{*++}&\frac{1}{\sqrt{2}}\Sigma^{*+}_c&\frac{1}{\sqrt{2}}\Xi^{*+}_c\\
\frac{1}{\sqrt{2}}\Sigma^{*+}_c&\Sigma_c^{*0}&\frac{1}{\sqrt{2}}\Xi^{*0}_c\\
\frac{1}{\sqrt{2}}\Xi^{*+}_c&\frac{1}{\sqrt{2}}\Xi^{*0}_c&\Omega^{*0}_c
\end{array}\right).\label{MBB}
\end{align}

The masses of  particles involved in the calculation are chosen as suggested
central values in the Review of  Particle Physics
(PDG)~\cite{Tanabashi:2018oca}. The mass of the  broad scalar $\sigma$ meson is
chosen as 500 MeV.  The  coupling constants involved are listed in
Table~\ref{coupling}.
\renewcommand{\tabcolsep}{0.16cm}
\renewcommand{\arraystretch}{1.5}
\begin{table}[h!]
\footnotesize
\caption{The coupling constants adopted in the
calculation, which are cited from the literature~\cite{Chen:2019asm,Liu:2011xc,Isola:2003fh,Falk:1992cx}. The $\lambda$, $\lambda_{S,I}$, and $f_\pi$ are in the unit of GeV$^{-1}$. Others are in units of $1$.
\label{coupling}}
\begin{tabular}{cccccccccccccccccc}\toprule[1pt]
$\beta$&$g$&$g_V$&$\lambda$ &$g_{s}$&$f_\pi$\\
0.9&0.59&5.9&0.56 &0.76 &0.132\\\hline
$\beta_S$&$\ell_S$&$g_1$&$\lambda_S$ &$\beta_B$&$\ell_B$ &$g_4$&$\lambda_I$\\
-1.74&6.2&-0.94&-3.31&$-\beta_S/2$&$-\ell_S/2$&$3g_1/{(2\sqrt{2})}$&$-\lambda_S/\sqrt{8}$ \\
\bottomrule[1pt]
\end{tabular}
\end{table}

\subsection{Lagrangians for light hadrons}

In the following we will present the Lagrangians for the vertex of the
constituent light hadrons $\bar{K}^{(*)}$ or $\Xi^{(*)}$ and the exchanged light hadrons $m$. The vertices
$\bar{K}^{(*)}\bar{K}^{(*)}m$ and $\Xi^{(*)}\Xi^{(*)}m$ can be related to vertices $\pi\pi m$, $\rho\rho m$, $\rho\omega m$, $NNm$, $\Delta\Delta m$, and $N\Delta$ under SU(3) flavor symmetry~\cite{deSwart:1963pdg,Ronchen:2012eg,Matsuyama:2006rp}.
 First, the Lagrangians for $\bar{K}^{(*)}\bar{K}^{(*)}m$ are shown as
\begin{eqnarray}
\mathcal{L}_{\bar{K}\bar{K}{V} }&=&ig_{\bar{K}\bar{K}V}\bar{K}^{\dag}
{V}^{\mu}\overleftrightarrow{\partial}_{\mu}\bar{K},   \nonumber\\
\mathcal{L}_{\bar{K}\bar{K}\sigma }&=&-g_{\bar{K}\bar{K}\sigma}\bar{K}^{\dag} \sigma\bar{K}, \nonumber\\
\mathcal{L}_{\bar{K}^*\bar{K}^*P}&=&g_{\bar{K}^*\bar{K}^*{P}}~\epsilon^{\mu\nu \alpha\beta}\partial_\mu \bar{K}^{*\dag}_\nu\partial_\alpha{P} \bar{K}^{*}_\beta~,\nonumber\\
\mathcal{L}_{\bar{K}^*\bar{K}^*{V}}&=&-i\frac{g_{\bar{K}^*\bar{K}^{*}{V}}}{2} ~ (\bar{K}^{*\mu\dag}{{V}}_{\mu\nu}\bar{K}^{*\nu}+\bar{K}^{*\dag}_{\mu\nu}{{V}}^{\mu}\bar{K}^{*\nu}
+\bar{K}^{*\mu\dag}{{V}}^{\nu}\bar{K}^{*}_{\nu\mu})~,\nonumber\\
\mathcal{L}_{\bar{K}^*\bar{K}^*\sigma}&=&g_{\bar{K}^*\bar{K}^{*}\sigma}\bar{K}^{*\mu\dag}\sigma \bar{K}^{*\mu}~,\nonumber\\
\mathcal{L}_{\bar{K}\bar{K}^*{P} }&=&ig_{\bar{K}\bar{K}^*{P}}\bar{K}_{\mu}^{*\dag}{P}\partial_{\mu}\bar{K}+h.c.~,\nonumber\\
\mathcal{L}_{\bar{K}\bar{K}^*{V}}&=&g_{\bar{K}\bar{K}^*{V}}~\epsilon^{\mu\nu \alpha\beta}\partial_\mu \bar{K}^{*\dag}_\nu\partial_\alpha{P} \bar{K}_\beta~,
\end{eqnarray}
where the ${P}$ and ${V}_\mu$ in the Lagrangians stand for the pseudoscalar meson
($\vec{\tau}\cdot\vec{\pi}$ or $\eta$) and vector meson
($\vec{\tau}\cdot\vec{\rho}_\mu$, $\omega_\mu$, or $\phi_\mu$), respectively; ${V}_{\mu\nu}=\partial_{\mu}{V}_{\nu}-\partial_{\nu}{V}_{\mu}$.
The  coupling constants can be obtained by the SU(3) relation and are listed in
Table~\ref{SUK}.
\renewcommand\tabcolsep{0.18cm}
\renewcommand{\arraystretch}{1.5}
\begin{table}[h!]
\footnotesize
\caption{The coupling constants determined with SU(3) symmetry. The values are  in the unit of GeV. The three basic constants are chosen as $g_{PPV}=3.02$, $g_{VVP}=5.6$ and $g_{VVV}=3.25$~\cite{Ronchen:2012eg}.\label{SUK}}
\begin{tabular}{lrrlrr}\toprule[1pt]
Coupl. 				&SU(3)	 relation	& Values			 				&Coupl.			                                  &SU(3) relation 		&Values\\ \hline
$g_{\bar{K}^*\bar{K}^*\pi}$       &$g_{VVP}$&$5.6$                           & $g_{\bar{K}^*\bar{K}^*\eta}$  &$-\sqrt{3}g_{VVP}$  &$-9.7$ \\
$g_{\bar{K}^*\bar{K}^*\rho}$	&$g_{VVV}$ & $3.25$                      &  $g_{\bar{K}^*\bar{K}^*\omega}$  & $-g_{VVV}$         & $-3.25$    \\
$g_{\bar{K}^*\bar{K}^*\phi}$&$-\sqrt{2}g_{VVV}$  & $-4.6$        &$g_{\bar{K}^*\bar{K}^*\sigma}$&$--$  & $3.65$\\
$g_{\bar{K}\bar{K}\rho}$                     &$g_{PPV}$           &$3.02$     &$g_{\bar{K}\bar{K}\omega}$    &$-g_{PPV}$              &$-3.02$ \\
$g_{\bar{K}\bar{K}\phi}$                &$-\sqrt{2}g_{PPV}$   &$-4.27$&$g_{\bar{K}\bar{K}\sigma}$&$--$  & $3.65$\\
$g_{\bar{K}\pi\bar{K}^*}$   &$-g_{PPV}$                  &$-3.02$            &$g_{\bar{K}\eta\bar{K}^*}$        & $\sqrt{3}g_{PPV}$   &$5.23$\\
$g_{\bar{K}^*\rho\bar{K}}$&$g_{VVP}$                  &$5.6$            &$g_{\bar{K}^*\omega\bar{K}}$              &$-g_{VVP}$&$-5.6$\\
$g_{\bar{K}^*\phi\bar{K}}$&$-\sqrt{2}g_{VVP}$ &$-7.92$                &&&\\
\bottomrule[1pt]\hline
\end{tabular}
\end{table}

The Lagrangians for vertices of light baryon coupling with a light exchanged mesons $\Xi^{(*)}\Xi^{(*)}m$  are written as
\begin{eqnarray}
\mathcal{L}_{\Xi\Xi {P}}&=&-\frac{g_{\Xi\Xi {P}}}{m_{{P}}} ~\bar{\Xi}~\gamma^5 \gamma^{\mu} \partial_{\mu} {P}~\Xi~,  \nonumber \\
\mathcal{L}_{\Xi\Xi{V}}&=& -~\bar{\Xi}~ [g_{\Xi\Xi {V}}\gamma^{\mu}-\frac{f_{\Xi\Xi {V}}}{2m_{\Xi}}\sigma^{\mu\nu}\partial_{\nu}]~ {V}_{\mu}\Xi,\nonumber\\
\mathcal{L}_{\Xi\Xi\sigma} &=&-g_{\Xi\Xi\sigma}~\bar{\Xi}~\sigma~\Xi~,%
\nonumber\\
\mathcal{L}_{\Xi^*\Xi^* {P}}&=&-\frac{g_{\Xi^*\Xi^* {P}}}{m_{{P}}} ~\bar{\Xi}^{*\alpha}~\gamma^5 \gamma^{\mu} \partial_{\mu} {P}~\Xi^*_\alpha~,\nonumber \\
  \mathcal{L}_{\Xi^*\Xi^*  {V}}&=& -~\bar{\Xi}^{*\alpha}~ [g_{\Xi^*\Xi^*{V}}\gamma^{\mu}-\frac{f_{\Xi^*\Xi^* {V}}}{2m_{\Xi^*}}\sigma^{\mu\nu}\partial_{\nu}] ~{V}_{\mu}\Xi^{*}_\alpha~,\nonumber\\
\mathcal{L}_{\Xi^*\Xi^*\sigma} &=&g_{\Xi^*\Xi^*\sigma}~\bar{\Xi}^{*\mu}~\sigma~\Xi^*_{\mu}~,
\nonumber\\
\mathcal{L}_{ \Xi\Xi^{*}{P}}&=&\frac{g_{\Xi\Xi^* {P}}}{m_{{P}}} \bar{\Xi}^{*\mu}~\partial_{\mu} {P}~\Xi +h.c.~ ,\nonumber \\
  \mathcal{L}_{\Xi\Xi^*{V}}&=& -i\frac{g_{\Xi\Xi^*{V}}}{m_{{V}}}\bar{\Xi}^{*\mu} \gamma^{5}\gamma^{\nu} {V}_{\mu\nu}\Xi~ +h.c. .
\end{eqnarray}
With the help of the SU(3) symmetry, the values of the coupling constants are given in Table~\ref{tab:coupl2}. Here, we choose $\alpha_{BBV}= 1.15$ as in Ref.~\cite{Ronchen:2012eg} based on the $NN\omega$ coupling constant given in
Refs.~\cite{Janssen:1996kx,Reuber:1993ip} differing from the standard value $\alpha_{BBV}= 1$, which introduces a small SU(3) symmetry
breaking in all vector-meson couplings. Furthermore, we adopt $f_{NN\omega} = 0 $ with  $f_{BBV}= g_{BBV}\kappa_{V}$,
$f_{DDV}= g_{DDV}\kappa_{V}$, and
$\kappa_{\rho}=6.1$.
\renewcommand\tabcolsep{0.16cm}
\renewcommand{\arraystretch}{1.5}
\begin{table}[h!]
\caption{The values of coupling constants and SU(3) relations. The values are in the unit of GeV. The basic constants  $g_{BBP}=0.989$,  $g_{BBV}=3.25$, $\alpha_{BBP}=0.4$, $\alpha_{BBV}=1.15$ and $g_{BB\sigma}=6.59$; $g_{DDP}=13.79$, $g_{DDV}=59.41$, $g_{BDP}=9.48$, and $g_{BDP}=71.69$~\cite{Ronchen:2012eg,Matsuyama:2006rp}.\label{tab:coupl2}}
\footnotesize
\begin{tabular}{lrrlrr}\toprule[1pt]
\hline
Coupl. 				&SU(3) relation		& Values			 				&Coupl.			&SU(3) relation 		&Values	 \\ \hline
$g_{\Xi\Xi\pi}$		&$(2\alpha-1)g_{BBP}$&$-0.20$&$g_{\Xi\Xi\eta}$& $-\frac{\sqrt{3}(1+2\alpha)}{3}g_{BBP}$	& $-1.03$\\
$g_{\Xi\Xi\rho}$		&$(2\alpha-1)g_{BBV}$&$4.23$&$g_{\Xi\Xi\omega}$& $(2\alpha-1)g_{BBV}$	& $4.23$\\
$g_{\Xi\Xi\sigma}$&$g_{BB\sigma}$                             &$6.59$&$g_{\Xi^*\Xi^*\sigma}$&$g_{BB\sigma}$              &$6.59$\\
$g_{\Xi^*\Xi^*\pi}$&$\frac{1}{4\sqrt{15}}g_{DDP}$&$0.89$&$g_{\Xi^*\Xi^*\eta}$&$-\frac{1}{4\sqrt{5}}g_{DDP}$& $-1.54$\\
$g_{\Xi^*\Xi^*\rho}$&$\frac{1}{4\sqrt{15}}g_{DDV}$&$3.84$&$g_{\Xi^*\Xi^*\omega}$&$\frac{1}{4\sqrt{15}}g_{DDV}$& $3.84$\\
$g_{\Xi\Xi^*\pi}$&$\frac{1}{2\sqrt{30}}g_{BDP}$&$0.87$&$g_{\Xi\Xi^*\eta}$&$-\frac{1}{2\sqrt{10}}g_{BDP}$& $-1.50$\\
$g_{\Xi\Xi^*\rho}$&$\frac{1}{2\sqrt{30}}g_{BDV}$&$6.54$&$g_{\Xi\Xi^*\omega}$&$-\frac{1}{2\sqrt{30}}g_{BDV}$& $-6.54$\\
$f_{\Xi\Xi\rho}$ &$\frac{1}{2}(f_{NN\omega}-f_{NN\rho})$ &-9.9&
$f_{\Xi\Xi\omega}$&$\frac{1}{2}(f_{NN\omega}-f_{NN\rho})$ &-9.9\\
$f_{\Xi^*\Xi^*\rho}$&$\frac{1}{4\sqrt{15}}f_{\Delta\Delta\rho}$ &29.4&
$f_{\Xi^*\Xi^*\omega}$&$\frac{1}{4\sqrt{15}}f_{\Delta\Delta\rho}$ &29.4\\
\bottomrule[1pt]\hline
\end{tabular}
\end{table}
\subsection{Potential kernels}
The isospin wave functions for the systems considered in the current work can be written as,
\begin{eqnarray}
  &| \Xi^{(*)}  D^{(*)} , I=0 \rangle=& -\frac{1}{\sqrt{2}} \Big|\Xi^{(*)0}D^{(*)0}+\Xi^-D^+\Big\rangle\, ,\nonumber\\
  &| \Xi^{(*)} D^{(*)} , I=1 \rangle=& -\frac{1}{\sqrt{2}} \Big|\Xi^{(*)0} D^{(*)0} - \Xi^{*-} D^+\Big\rangle\,,\nonumber\\
  &| \Xi_c ^{(*,')}\bar{K}^{(*)}, I=0 \rangle=& -\frac{1}{\sqrt{2}} \Big|\Xi_c^{(*,')+} K^{(*)-} + \Xi_c^{(*)0} \bar{K}^{(*)0} \Big\rangle\, ,\nonumber\\
   &| \Xi_c^{(*,')} \bar{K}^{(*)}, I=1 \rangle=& -\frac{1}{\sqrt{2}} \Big|\Xi_c^{(*,')+} K^{(*)-} - \Xi_c^{(*)0}  \bar{K}^{(*)0} \Big\rangle\, ,\nonumber\\
  &|\Omega_c ^{(*)}\omega(\eta), I=0 \rangle=&  \Big|\Omega_c ^{(*)0}\omega(\eta) \Big\rangle\, ,\nonumber\\
  &|\Omega_c ^{(*)}\rho(\pi), I=1 \rangle=&  \Big|\Omega_c ^{(*)0}\rho^0(\pi^0)\Big\rangle\, .
  \end{eqnarray}
Here, the isospin multiplets are defined as
\begin{eqnarray}
D^{(*)} = \myvec{D^{(*)+}\\-D^{(*)0}},\,\, \Xi^{(*)} = \myvec{\Xi^{(*) 0}\\ \Xi{^{(*) -}}};
 \vspace{15pt}\nonumber
\end{eqnarray}
\begin{eqnarray}
\bar{K}^* = \myvec{\bar{K}^{(*)0}\\-K^{(*)-}},\,\,\ \Xi^{(*)}_{c} = \myvec{\Xi_{c}^{(*)+}\\ \Xi_c^{(*)0}},\,\, \ \Xi'_{c} = \myvec{\Xi_{c}^{'+}\\ \Xi_{c}^{'0}}.
 \vspace{15pt}
\end{eqnarray}

With the vertices obtained from  the above Lagrangians, the potential kernels of
coupled-channel interactions can be constructed easily with the help of the
standard Feynman rules. In this work, following the method in
Refs.~\cite{He:2019rva,Zhu:2021lhd,Zhu:2020vto}, we input vertices $\Gamma$ and
propagators $P$  into  the code directly. The potential can be written as
\begin{equation}%
{\cal V}_{\mathbb{P},\sigma}=I_{\mathbb{P},\sigma}\Gamma_1\Gamma_2 P_{\mathbb{P},\sigma}f(q^2),\ \
{\cal V}_{\mathbb{V}}=I_\mathbb{V}\Gamma_{1\mu}\Gamma_{2\nu}  P^{\mu\nu}_{\mathbb{V}}f(q^2).\label{V}
\end{equation}
The propagators are defined as usual as
\begin{equation}%
P_{\mathbb{P},\sigma}= \frac{i}{q^2-m_{\mathbb{P},\sigma}^2},\ \
P^{\mu\nu}_\mathbb{V}=i\frac{-g^{\mu\nu}+q^\mu q^\nu/m^2_{\mathbb{V}}}{q^2-m_\mathbb{V}^2},
\end{equation}
The form factor $f(q^2)$ is adopted to compensate the off-shell effect of
exchanged meson as $f(q^2)=e^{-(m_e^2-q^2)^2/\Lambda_e^2}$ with $m_e$ and $q$ being the mass and momentum of the exchanged meson.
The flavor factor $I_{ex}$ can be obtained with the wave functions and Lagrangians as listed in Table~\ref{flavor factor}.

The potential  kernel obtained in Eq.~(\ref{V}) can be projected into fixed spin-parity by partial-wave decomposition as
\begin{align}
{\cal V}_{\lambda'\lambda}^{J^P}({\rm p}',{\rm p})
&=2\pi\int d\cos\theta
~[d^{J}_{\lambda\lambda'}(\theta)
{\cal V}_{\lambda'\lambda}({\bm p}',{\bm p})\nonumber\\
&+\eta d^{J}_{-\lambda\lambda'}(\theta)
{\cal V}_{\lambda'-\lambda}({\bm p}',{\bm p})],
\end{align}
where $\eta=PP_1P_2(-1)^{J-J_1-J_2}$ with $P$ and $J$ being parity and spin for the system. The initial and final relative momenta are chosen as ${\bm p}=(0,0,{\rm p})$  and ${\bm p}'=({\rm p}'\sin\theta,0,{\rm p}'\cos\theta)$. The $d^J_{\lambda\lambda'}(\theta)$ is the Wigner $d$ matrix.

\renewcommand\tabcolsep{0.21cm}
\renewcommand{\arraystretch}{1.45}
\begin{table}[h!]
\footnotesize
 \centering
 \caption{The flavor factor ${I}_{ex}$ for a system with certain isospin and  meson exchange. The vertices for three pseudoscalar mesons should be forbidden.  \label{flavor factor}}
\begin{tabular}{ccccccc}\toprule[1pt]
  Transition&\multicolumn{4}{c}{$\Xi^{(*)}D^{(*)}\to\Xi^{(*)}D^{(*)}  $} \\
  ${I}_{ex}$& $I_\pi$& $I_\eta$&$I_\rho$&$I_\omega$ &$I_\phi$ &$I_\sigma$ \\
 $I=0$&$3/\sqrt2$ &$1/\sqrt6$ &$3/\sqrt2$ & $1/\sqrt2$&$--$&$1$\\
 $I=1$&$-1/\sqrt2$ &$1/\sqrt6$ &$-1/\sqrt2$ & $1/\sqrt2$&$--$&$1$\\
 Transition&\multicolumn{4}{c}{$\Xi^{(*,')}_c\bar{K}^{(*)} \to\Xi^{(*,')}_c\bar{K}^{(*)}  $} \\
  ${I}_{ex}$& $I_\pi$& $I_\eta$&$I_\rho$&$I_\omega$ &$I_\phi$ &$I_\sigma$ \\
 $I=0$&$-3/2\sqrt{2}$ &$-1/2\sqrt{6}$ &$-3/2\sqrt{2}$ & $1/2\sqrt{2}$&$1/2$&$1$\\
 $I=1$&$1/2\sqrt{2}$ &$-1/2\sqrt{6}$ &$1/2\sqrt{2}$ & $1/2\sqrt{2}$&$1/2$&$1$\\
  Transition&\multicolumn{4}{c}{$\Xi_c\bar{K}^{(*)} \to\Xi_c\bar{K}^{(*)} $} \\
   ${I}_{ex}$& $I_\pi$& $I_\eta$&$I_\rho$&$I_\omega$ &$I_\phi$ &$I_\sigma$ \\
 $I=0$&$--$ &$--$ &$-3/\sqrt2$ & $1/\sqrt2$&$1$&$2$\\
 $I=1$&$--$ &$--$ &$1/\sqrt2$ & $1/\sqrt2$&$1$&$2$\\
Transition&\multicolumn{4}{c}{$\Xi^{(*,')}_c\bar{K}^{(*)}\to\Xi_c \bar{K}^{(*)} $} \\
 ${I}_{ex}$& $I_\pi$& $I_\eta$&$I_\rho$&$I_\omega$ &$I_\phi$ &$I_\sigma$ \\
 $I=0$&$-3/2$ &$\sqrt3/2$ &$-3/2$ & $1/2$&$1$&$--$\\
 $I=1$&$1/2$ &$\sqrt3/2$ &$1/2$ & $1/2$&$1$&$--$\\
\hline
 Transition&\multicolumn{4}{c}{$\Omega^{(*)}_c\eta/\omega\to\Xi^{(*,')}_c/\bar{K}^{(*)} $} \\
 $I_{ex}$&  $I_K$&$I_{K^*}$&$ $ &$ $ &$ $ &\\
 $I=0$&$-1$ &$-1$ & & &&\\
 Transition&\multicolumn{4}{c}{$\Omega^{(*)}_c\pi/\rho\to\Xi^{*,'}_c\bar{K}^{(*)} $} \\
 $I_{ex}$&  $I_K$&$I_{K^*}$&$ $ &$ $ &$ $ &\\
 $I=1$&$1$ &$1$ & & &&\\
 Transition&\multicolumn{4}{c}{$\Omega^{(*)}_c\eta/\omega\to\Xi_c\bar{K}^{(*)}$} \\
 $I_{ex}$&  $I_K$&$I_{K^*}$&$ $ &$ $ &$ $ &\\
 $I=0$&$-\sqrt2$&$-\sqrt2$& & &&\\
 Transition&\multicolumn{4}{c}{$\Omega^{(*)}_c\pi/\rho\to\Xi_c\bar{K}^{(*)}$} \\
$I_{ex}$&  $I_K$&$I_{K^*}$&$ $ &$ $ &$ $ &\\
 $I=1$&$\sqrt2$ &$\sqrt2$ & & &&\\
\bottomrule[1pt]
\end{tabular}
\end{table}

To obtain the scattering amplitude, such partial-wave potential kernel can be inserted into the partial-wave quasipotential Bethe-Salpeter equation as~\cite{He:2014nya,He:2015mja,He:2012zd,He:2015yva,He:2017aps}
\begin{align}
  i{\cal M}^{J^P}_{\lambda'\lambda}({\rm p}',{\rm p})
  &=i{\cal V}^{J^P}_{\lambda',\lambda}({\rm p}',{\rm
  p})+\sum_{\lambda''\geq0}\int\frac{{\rm
  p}''^2d{\rm p}''}{(2\pi)^3}\nonumber\\
  &\cdot
  i{\cal V}^{J^P}_{\lambda'\lambda''}({\rm p}',{\rm p}'')
  G_0({\rm p}'')i{\cal M}^{J^P}_{\lambda''\lambda}({\rm p}'',{\rm
  p}),\quad\quad \label{Eq: BS_PWA}
  \end{align}
The reduced propagator $G_0({\rm p}'')$ under quasipotential approximation has a
form of $G_0({\rm p}'')=\delta^+(p''^{~2}_h-m_h^{2})/(p''^{~2}_l-m_l^{2})$ with
$p''_{h,l}$ and $m_{h,l}$ being the momenta and masses of heavy or light
constituent particles.  Since in the current approach, one of the constituent
particles is put on shell while another is put off shell, an  exponential regularization is also introduced
as $G_0({\rm p}'')\to G_0({\rm
p}'')\left[e^{-(p''^2_l-m_l^2)^2/\Lambda_r^4}\right]^2$ with $\Lambda_r$ being a
cutoff~\cite{He:2015mja}. To make it more in line with the values of different exchange mesons, a parameterization on the cutoff is performed as
$\Lambda_e=\Lambda_r=m_e+\alpha~0.22$~GeV with $m_e$ being the mass of the
exchange meson.

The partial-wave Bethe-Salpeter equation can be transformed into a matrix
equation as $M=V+VG_0M$ by Gauss discretizing, with which the scattering
amplitude can be obtained as  $M(z)=V(z)/(1-V(z)G_0(z))$. The molecular state
can be found at the pole of scattering amplitude in the complex energy plane when
its denominator $|1-V(z)G_0(z)|=0$. The real and imaginary parts of the pole
position $z=W+i\Gamma/2$ are energy and half of the decay width, respectively.

\section{Results with single-channel calculation \label{sec3}}

Because of the complexity of the coupled-channel results,  the results with
single-channel calculation will be first presented in this section, which can
provide an overall picture of the molecular states produced from the interactions
considered.

\subsection{Category I: Interaction $\Xi^{(*)}D^{(*)}$}\label{sec3.1}

\begin{figure}[h!]
  \centering
\includegraphics[scale=1.6,bb=45 300 150 480 ]{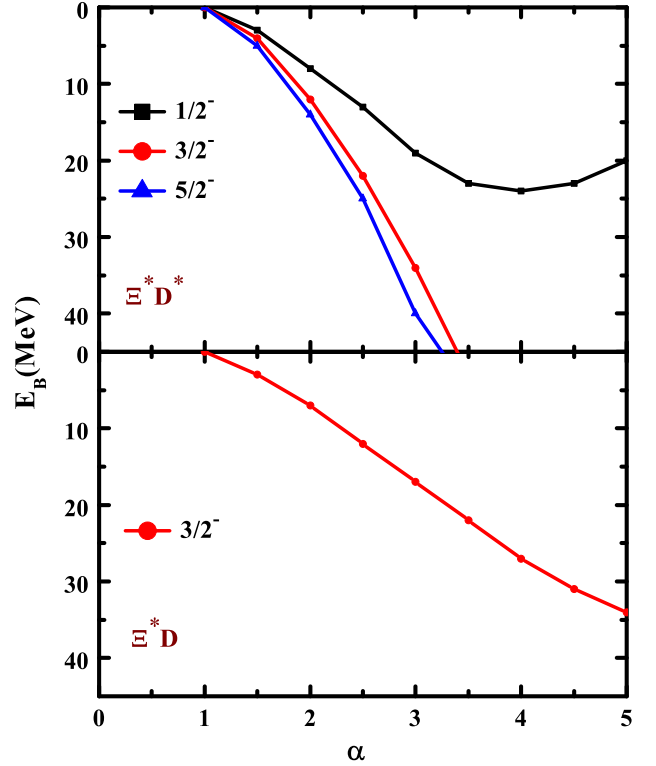}
\caption{The binding energies  for the isoscalar bound states from interaction $\Xi^{(*)}D^{(*)}$ with variation of $\alpha$ with single-channel calculation. The upper and lower panels are for interactions $\Xi^*D^*$ and  $\Xi^*D$, respectively.}\label{1}
\end{figure}

In Fig.~\ref{1}, the binding energies of all bound states produced from
interaction $\Xi^{(*)}D^{(*)}$ are presented with the variation of the $\alpha$, which is the only variable parameter in the current model, and absorbs
the model uncertainties. Empirically, its value is chosen in a reasonable range
from 0 to 5.  Since the molecular state is defined as a shallow bound state,
in the current work, we only keep the results with binding energy lower than $45$
MeV or smaller.

The single-channel calculation suggests that only the isoscalar bound state can
be produced from the interactions. No isovector state is found in the considered
range of the parameter. Three isoscalar $\Xi^*D^*$ states appear at almost the
same value of $\alpha$ about 1 with spin parities  $J^P=1/2^-$, $3/2^-$, and
$5/2^-$, which are all possible spin parities in the $S$-wave.  The binding energies
of all three states gradually increase with the increase of the $\alpha$. When
$\alpha$ increases to 4 or larger, the repulsion  from $\pi$ exchange increases faster than the attraction, which makes the state with
$1/2^-$ shallower.

Different from the $\Xi^*D^*$ interaction, only one molecular
state can be produced from the $\Xi^* D$ interaction with spin parity
$3/2^-$.  For two lower interactions $\Xi D^*$ and $\Xi D$, no bound
state is found near their thresholds for all quantum numbers considered in the
current work as listed in Table~\ref{T1}.

\subsection{Category II: Interaction $\Xi_c^{(*,')}\bar{K}^{(*)}$}\label{sec3.2}

\begin{figure}[h!]
  \centering
  \includegraphics[scale=1.05,bb=20 215 265 480]{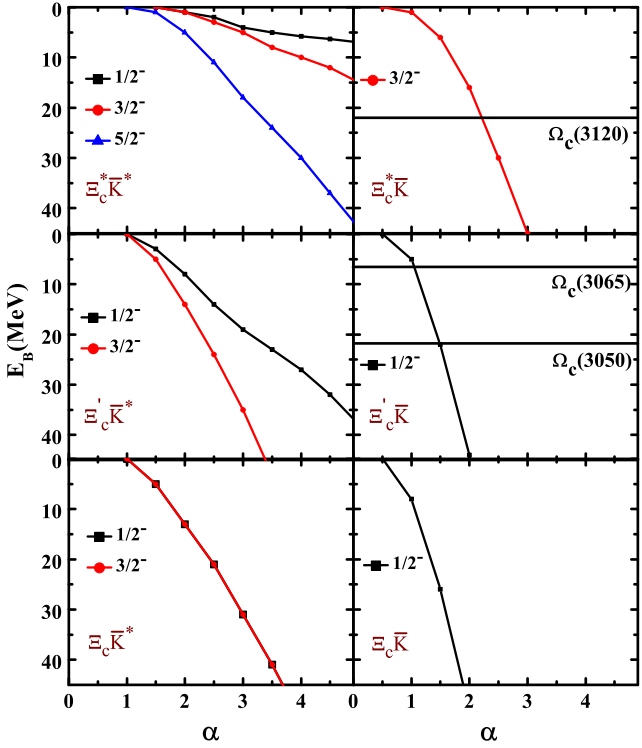}\\
  \caption{Binding energies of isoscalar bound states from the interactions $\Xi_c^{(*,')}\bar{K}^{(*)}$ with the variation of $\alpha$ with single-channel calculation. The horizontal lines are for the experimental values of the masses of corresponding states~\cite{LHCb:2021ptx}. The very small uncertainties of masses, only a few tenths of MeV, are not plotted.}\label{2}
\end{figure}

Six interactions in category II, $\Xi_c^*\bar{K}^*$, $\Xi'_c\bar{K}^* $,
 $\Xi_c\bar{K}^*$, $\Xi_c^*\bar{K}$, $\Xi'_c\bar{K} $, and $\Xi_c\bar{K}$, involve in the single-channel calculation. We
 first present the binding energies of isoscalar states with $I=0$ from these interactions in Fig.~\ref{2}.

The $\Xi_c^*\bar{K}^*$ interaction is found attractive and produces three
isoscalar bound states with spin parities $1/2^-$, $3/2^-$, and $5/2^-$, which
all appear at a value of $\alpha$ of about 1. Their binding energies increase with
the increase of the parameter $\alpha$. The binding energies of states with
larger spin increase more rapidly. Two bound states are produced from the
$\Xi'_c\bar{K}^*$ interaction with $1/2^-$ and $3/2^-$. Large binding energy is
also found for large spin parity $3/2^-$. The bound states from interaction
$\Xi_c\bar{K}^*$ with $1/2^-$ and $3/2^-$ are almost degenerate. It is from
small differences of the vector exchange for different spin parities and the
absence of pseudoscalar exchange as shown in Table~\ref{flavor factor}. A very
large $\alpha$ value is required to provide the obvious difference in binding energy.

As shown in the panels in the right of Fig.~\ref{2},  only one bound state is
produced for each of the interactions, $\Xi_c^*\bar{K}$, $\Xi'_c\bar{K}$, or
$\Xi_c\bar{K}$, and all of them appear at a value of $\alpha$ of about 0.5. The
curve of the binding energy of $\Xi^*_c\bar{K}$ state with $3/2^-$ is compared
with the experimental mass of the $\Omega_c(3120)$ as a horizontal line. The
cross can be found at an $\alpha$ value of about $2.2$. The bound state from
$\Xi'_c\bar{K}$ interaction also appears at a value of $\alpha$ of about 0.5. To
reach the experimental  mass of state $\Omega(3065)$ or
$\Omega(3050)$, an $\alpha$  value about 1.0 or 1.5 is required for isoscalar
state $\Xi'_c\bar{K}$ with $1/2^-$, respectively.

\begin{figure}[h!]
  \centering
  \includegraphics[scale=1.05,bb=20 215 265 480]{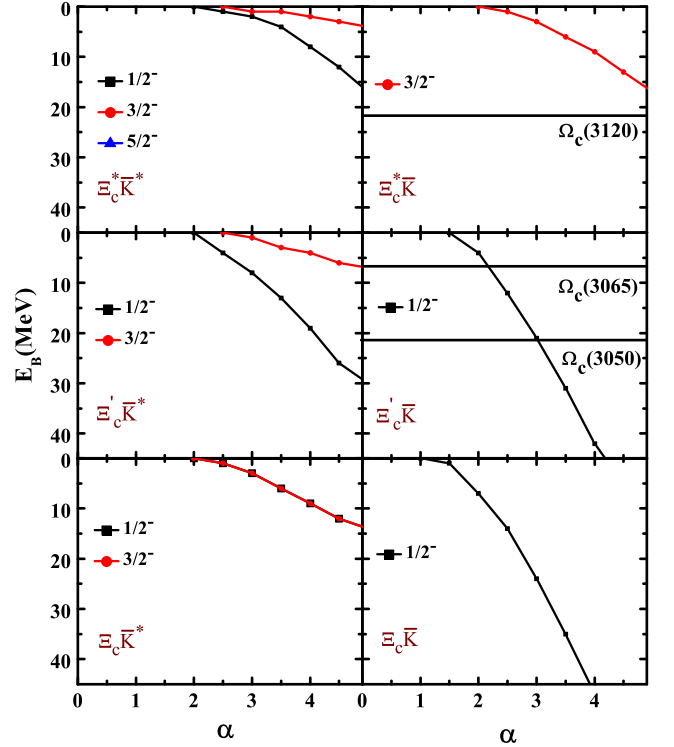}\\
  \caption{The binding energies of isovector bound states from the interaction $\Xi_c^{(*,')}\bar{K}^{(*)}$ with the variation of $\alpha$ with single-channel calculation. The horizontal lines are for the experimental values of the masses of corresponding states~\cite{LHCb:2021ptx}.}\label{3}
\end{figure}

In Fig.~\ref{3}, isovector bound states from the interactions
$\Xi_c^{(*,')}\bar{K}^{(*)}$ are listed. Generally, one can find that larger
$\alpha$ is required to form an isovector bound state than isoscalar states. It
is mainly due to the different flavor signs for the $\rho$ meson exchange in
Table~\ref{flavor factor}, which means attraction and repulsion for isoscalar
and isovector interaction, respectively.

For three interactions with the $\bar{K}^*$ meson,  $\Xi_c^*\bar{K}^*$,
$\Xi'_c\bar{K}^*$, or $\Xi_c\bar{K}^*$, two bound states with spin parities
$1/2^-$ and $3/2^-$ appear at an  $\alpha$ value of about 2, which is much larger
than these for isoscalar states. The binding energy of the state with spin parity
$3/2^-$ becomes smaller than the states with $1/2^-$ from interactions
$\Xi_c^*\bar{K}^*$ and $\Xi'_c \bar{K}^*$.  The isovector $\Xi_c^*\bar{K}^*$
state  with $5/2^-$ is attractive but too weak to produce a bound state in the 
considered range of parameter $\alpha$. Two $\Xi_c \bar{K}^*$ states with spin
parities $1/2^-$ and $3/2^-$ are also almost degenerate as in the isoscalar case.

For three interactions with a $\bar{K}$ meson involved,  only one bound state can be
produced as in isoscalar cases.  The state with spin parity $3/2^-$ from
interaction $\Xi_c^*\bar{K}$ appears at a value of $\alpha$ of about 2 and reaches a
binding energy of about $16$~MeV at the largest $\alpha$ value we considered.
Compared with the experimental mass of the $\Omega_c(3120)$, too large $\alpha$ of an
value is required to assign the isovector $\Xi_c^*\bar{K}$ state with $3/2^-$ as
$\Omega_c(3120)$.  For the isovector state $\Xi'_c\bar{K}$ with spin parity
$1/2^-$, a binding energy of about 7~MeV can be reached at $\alpha$ of about $2.2$,
which is close to mass of the states $\Omega_c(3065)$ and $\Omega_c(3050)$.

\section{Results with coupled-channel calculation}\label{sec4}

In the previous section, many bound states are found to  be produced from the
considered interactions with single-channel calculation in reasonable range of
$\alpha$. In this section, the coupled-channel effects will be introduced to
study the variation of the bound states obtained in the previous section and
couplings of the molecular states to relevant channels.

\subsection{Coupled-channel results for category I}

In Table~\ref{DDD}, coupled-channel results with all four channels are listed in
the third column with the label ``$cc$". Since the poles are found near the
corresponding thresholds, we still present position as $M_{th}-z$ with $M_{th}$
being the corresponding threshold as in the single-channel case in the previous
section for comparison. Besides, in the fourth to seventh columns, the results
for couplings to the channels as labeled are presented with two-channel
calculations.  The column without an imaginary part is the single-channel results
for reference. The couplings of channels above thresholds cannot produce width
as expected, which is reflected by a zero imaginary part ``$0i$".

\renewcommand\tabcolsep{0.2cm}
\renewcommand{\arraystretch}{1.5}
\begin{table}[h!]    
\footnotesize
\caption{The masses and widths of molecular states from interactions in category I at
different values of $\alpha$ in the unit of GeV. The values of the complex
position mean the mass of the corresponding threshold subtracted by the position of a
pole, $M_{th}-z$, in the unit of MeV. The imaginary part of some poles is shown
as 0.0, which means too small  a value under the current precision chosen. The
explicit explanation can be found in the text.\label{DDD}}
\begin{tabular}{c|cr|rrrr}\toprule[1pt]
\multicolumn{7}{c}{Poles near the $\Xi^* D^*~$ threshold with $M_{th}=3541.8$~MeV}\\
\hline
$IJ^P$&$\alpha$ &$cc$ &{$\Xi^* D^*$ }  &{$ \Xi^* D$ } &{$\Xi D^*$ }    &{$\Xi D$  }   \\ \hline
                                           &   $1.5$  &$3+1.1i$    &$3$    &$ 4+1.7i$         &$4+0.2i$      &$4+0.0i$    \\
$0\frac{1}{2}^- $   &  $2.0$   &$11+1.9i$   &$8$    &$ 13+1.5i$       &$10+0.4i$    &$9+0.1i$    \\
                                           &  $2.5$  &$19+5.1i$   &$14$   &$24+2.4i$           &$16+0.8i$    &$14+0.8i$   \\
                                        &    $3.0$   &$--$  &$19$   &$ 36+3.1i $         &$23+0.9i$    &$20+2.9i$   \\
\hline
                                 &    $1.5$     &$4+0.0i$   &$4$       &$5+0.1i$         &$5+0.1i$      &$5+0.0i$    \\
$0\frac{3}{2}^-$  &$2.0$   &$14+1.0i$   &$12$    &$12+0.2i$     &$14+0.7i$    &$13+0.1i$    \\
                                  &   $2.5$   &$26+2.7i$  &$22$       &$23+0.4i$       &$25+1.9i$    &$23+0.3i$   \\
                                  &   $3.0$   &$41+3.5i$  &$34$        &$36+1.0i$       &$38+3.1i$    &$35+0.7i$   \\
\hline
                                                     &$1.5$    &$5+0.8i$    &$5$       &$5+0.8i$         &$6+0.0i$      &$6+0.0i$    \\
    $ 0\frac{5}{2}^- $    & $2.0$ &$11+2.3i$     &$14$    &$11+1.5i$     &$15+0.1i$    &$14+0.2i$    \\
                                               & $2.5$   &$22+1.6i$  &$25$       &$22+1.0i$       &$26+0.3i$    &$26+0.8i$   \\
                                                    &$3.0$   &$37+1.7i$  &$40$        &$38+0.6i$       &$41+0.4i$    &$41+1.5i$   \\
\hline
\multicolumn{7}{c}{Poles near the $\Xi^* D$ thresholds with $M_{th}=3400.6$~MeV}\\\hline
$IJ^P$&$\alpha$  &$cc$&{$\Xi^* D^*$ } &{$ \Xi^* D$ }  &{$\Xi D^*$ }   &{$\Xi D$}   \\ \hline
                                              &  $1.5$     &$4+0.7i$   &$4+0i$       &$3$         &$4+0.6i$      &$3+0.0i$    \\
   $0\frac{3}{2}^-$    &   $2.0$    &$10+1.1i$      &$9+0i$         &$8$       &$9+1.1i$        &$8+0.0i$    \\
                                             &    $2.5$    &$18+1.4i$   &$16+0i$       &$13$       &$15+1.4i$    &$13+0.0i$   \\
                                               &     $3.0$   &$30+1.5i$  &$26+0i$      &$18$        &$22+1.5i$      &$18+0.1i$   \\
\hline
\bottomrule[1pt]
\end{tabular}
\end{table}

For isoscalar $\Xi^*D^*$ states with spin parities $1/2^-$ and $3/2^-$, the
introduction of three coupled channels leads to a smaller mass as well as a
nonzero width.  The pole for the $\Xi^*D^*$ state with $1/2^-$ gradually becomes
too dim to be seen in the complex plane with the increase of $\alpha$.  After
inclusion of  coupled-channel effects, the isoscalar $\Xi^* D^*$ state with
$5/2^-$ becomes shallower than that with $3/2^-$, and its total decay width first
increases and then decreases.  According to the imaginary parts in the last
three columns, isoscalar state  $\Xi^* D^*$  with $1/2^-$ has the strongest
coupling to the $\Xi^* D$ channel, while state with $3/2^-$ state prefers the $\Xi D^*$
channel. For the $\Xi^* D^*$ state with $5/2^-$, the dominant channel shifts from $\Xi^*
D$ to $\Xi D$ with the increase of $\alpha$.  For $\Xi^* D$ interaction, from
which only one state can be formed in the $S$-wave with $3/2^-$, the coupled-channel
effect obviously elevates its binding energy.  It can hardly couple to the $\Xi D$
channel but strongly couple to the $\Xi D^*$ channel.

\subsection{ Coupled-channel results for category II}

In single-channel calculation, the bound states from the interactions in
category II are presented, and their relations to the experimentally observed
states $\Omega_c(3120)$, $\Omega_c(3065)$, and $\Omega_c(3050)$ are also
discussed. In the following, the coupled-channel effects will be investigated.
In the single-channel calculation, the $\Omega^{(*)}_c(\pi/\eta/\rho/\omega)$
interactions are not included due to  the absence of  exchange of light
mesons. Such interactions should be included in the following coupled-channel
calculation due to  the existence of  $K$ or $K^*$  exchange between these
interactions.

{\noindent $\bullet$ \it Poles near the $\Xi^*_c \bar{K}$ threshold}

In the first part of Table~\ref{c45}, we present the coupled-channel results
near the $\Xi^*_c \bar{K}$ threshold, which is close to the state
$\Omega_c(3120)$. Here, as well as for other thresholds in the table, the couplings
to the channels with very high masses are extremely weak. Hence, though the overall coupled-channel results with all ten channels are listed in
the ``$cc$" column, only four important  two-channel results are listed in the table.

\renewcommand\tabcolsep{0.096cm}
\renewcommand{\arraystretch}{1.5}
\begin{table}[h!]    
\footnotesize
\caption{The masses and widths of molecular states from interactions in category
II near thresholds with a $\bar{K}$ meson at different values of $\alpha$ in the unit of GeV. Other notations are the same as Table~\ref{DDD}. }\label{c45}
\begin{tabular}{c|cr|rrrrr}\toprule[1pt]
\multicolumn{8}{c}{Poles near the $\Xi^*_c \bar{K}$ threshold with  $M_{th}=3142.0$~MeV}\\
\hline
$IJ^P$&$\alpha$     &$ cc$  &{$ \Omega^*_c \eta$ }         &{$\Omega_c \eta$ } &{$\Xi^*_c \bar{K}$ } &{$\Xi'_c \bar{K}$  }   &{$\Xi_c \bar{K}$\ \  } \\ \hline
                                      &$1.0$       &$1+0.0i$   &$1+0i$     &$1+0i$    &$1$        &$1+0.00i$     &$1+0.00i$  \\
                                      &$1.5$       &$8+0.02i$   &$ 8+0i$   &$6+0i$   &$6$         &$6+0.00i$      &$6+0.01i$    \\
$0\frac{3}{2}^-$                      &$2.0$       &$21+0.05i$  &$20+0i$   &$16+0i$ &$16$     &$16+0.00i$    &$16+0.02i$    \\
                                      &$2.5$       &$40+0.12i$  &$36+0i$     &$30+0i$&$30$       &$30+0.00i$    &$30+0.03i$   \\
\hline
$IJ^P$&$\alpha$       &$ cc$  &{$\Xi^*_c \bar{K}$ }        &{$\Xi'_c \bar{K}$  }
&{$\Xi_c \bar{K}$ }&{$ \Omega^*_c \pi$ }         &{$\Omega_c \pi$ }  \\ \hline
                                      &$2.0$    &$1+0.4i$  &$0$     &$0+0.0i$    &$0+0.0i$  &$ 1+0.3i$   &$0+0.0i$  \\
 $1\frac{3}{2}^-$                     &$2.5$   &$2+2.0i$ &$1$       &$1+0.0i$    &$1+0.0i$&$2+1.5i$     &$1+0.0i$   \\
                                      &$3.0$    &$6+4.1i$ &$3$     &$3+0.0i$     &$3+0.0i$ &$5+3.6i$     &$3+0.0i$ \\
                                      &$3.5$    &$13+6.6i$  &$6$   &$6+0.0i$      &$6+0.0i$    &$9+4.5i$   &$6+0.0i$\\
\hline
\multicolumn{8}{c}{Poles near the $\Xi'_c \bar{K}$ threshold $M_{th}=3072.5$~MeV}\\
\hline
$IJ^P$&$\alpha$      &$ cc$ &{$ \Omega^*_c \eta$ }  &{$\Omega_c \eta$ } &{$\Xi^*_c \bar{K}$ }        &{$\Xi'_c \bar{K}$  }   &{$\Xi_c \bar{K}$ } \\ \hline
                                  &$0.7$        &$2+0.0i$     &$0+0i$    &$0+0i$     &$1+0i$         &$0$     &$0+0.0i$  \\
                $0\frac{1}{2}^-$ &$1.0$        &$13+0.0i$     &$5+0i$    &$7+0i$     &$10+0i$ &$5$     &$5+0.0i$  \\
                                 &$1.5$      &$41+0.0i$   &$22+0i$    &$26+0i$ &$34+0i$ &$22$     &$22+0.0i$    \\
                                 &$2.0$      &$76+0.0i$    &$44+0i$   &$50+0i$  &$64+0i$ &$44$   &$44+0.0i$    \\
\hline
$IJ^P$&$\alpha$  &$ cc$  &$\Xi^*_c \bar{K}$     &$\Xi'_c \bar{K}$  &$\Xi_c \bar{K}$ &$ \Omega^*_c \pi$        &$\Omega_c \pi$   \\ \hline
                                     &$1.5$       &$4+3.8i$    &$2+0i $   &$0$ &$0+0.0i $  &$0+0.0i$   &$ 0+0.0i$    \\
$1\frac{1}{2}^-$                     &$2.0$       &$16+9.0i$    &$11+0i $   &$4$ &$4+0.0i $  &$4+0.7i$   &$ 6+5.7i$    \\
                                     &$2.5$        &$36+14.1i$  &$24+0i$  &$12$  &$12+0.0i $&$12+1.2i$    &$ 16+8.3i$   \\
                                      &$3.0$      &$62+20.3i$   &$39+0i$   &$21$  &$ 21+0.0i$ &$ 22+1.8i$    &$ 29+12.2i$   \\
\hline
\multicolumn{8}{c}{Poles near the $\Xi_c \bar{K}$ threshold with $M_{th}=2965.1$~MeV}\\
\hline
$IJ^P$&$\alpha$      &$ cc$ &{$ \Omega^*_c \eta$ }  &{$\Omega_c \eta$ } &{$\Xi^*_c \bar{K}$ }        &{$\Xi'_c \bar{K}$  }   &{$\Xi_c \bar{K}$ } \\ \hline
                                         &$0.5$        &$0+0.0i$     &$0+0i$    &$0+0i$     &$0+0i$       &$0+0i$  &$0$     \\
    $0\frac{1}{2}^-$ &$1.0$        &$8+0.0i$     &$8+0i$    &$8+0i$     &$8+0i$ &$8+0i$  &$8$     \\
                                         &$1.5$      &$26+0.0i$   &$26+0i$    &$26+0i$ &$26+0i$ &$26+0i$&$26$     \\
                                      &$2.0$      &$50+0.0i$    &$50+0i$   &$50+0i$  &$50+0i$&$50+0i$ &$50$   \\
\hline
$IJ^P$&$\alpha$  &$ cc$  &$\Xi^*_c \bar{K}$     &$\Xi'_c \bar{K}$  &$\Xi_c \bar{K}$ &$ \Omega^*_c \pi$        &$\Omega_c \pi$   \\ \hline
                                     &$1.5$       &$1+0.0i$    &$ 1+0i$   &$1+0i $ &$1$  &$1+0.0i$   &$ 1+0.0i$    \\
$1\frac{1}{2}^-$ &$2.0$       &$7+0.0i$    &$ 7+0i$    &$7+0i$  &$7$ &$ 7+0.0i$    &$ 7+0.0i$    \\
                                     &$2.5$        &$14+0.0i$ &$ 14+0i$ &$ 14+0i$&$14$ &$14+0.0i$  &$ 14+0.0i$   \\
                                      &$3.0$      &$24+0.0i$  &$ 24+0i$  &$ 24+0i$&$24$ &$ 24+0.0i$ &$ 24+0.0i$   \\
\hline
\bottomrule[1pt]
\end{tabular}
\end{table}

According to two-channel calculations, the pole of  isoscalar state with spin
parity $3/2^-$ near the $\Xi^*_c \bar{K}$ threshold is almost motionless after
inclusion of channels below the threshold, $\Xi'_c\bar{K}$, $\Xi_c\bar{K}$, and
the channels above the threshold, $\Omega_c\eta$. After inclusion
of all channels, the mass of this state becomes a little larger, which leads to
a smaller parameter $\alpha$, about $2.0$, required to reproduce the
experimental mass of the $\Omega_c(3120)$ than in single-channel calculation.
The variation of the mass is mainly from inclusion of the $\Omega_c^* \eta$
channel. The partial width to the $\Xi_c\bar{K}$ channel is small. If a
higher precision is adopted, a value of 0.06~MeV can be found at an $\alpha$ of
2.5, which is the main decay channel based on the current results.  The total
width with all channels considered is about 0.2~MeV.  If we vary  $\alpha$ to 3,
even a width with all channels of about $1$~MeV can be reached, which is close
to the central value of the experimental value as $\Gamma_{\Omega_c(3120)}=1.1 \pm
0.8 \pm0.4$ MeV.

The results for the  isovector $\Xi^*_c \bar{K}$ state with $3/2^-$ are also
presented in Table~\ref{c45}. In this case, the $\Omega_c^* \eta$ and $\Omega_c
\eta$ channels are replaced by $\Omega_c^* \pi$ and $\Omega_c \pi$ channels,
which are both under the threshold.  The $\Omega_c^*\pi$ channel is the
dominant decay channel, which gives a width at an order of magnitude of MeV.

{\noindent $\bullet$ \it Poles near the $\Xi'_c \bar{K}$ threshold}

In the second part of Table~\ref{c45}, we present the results for the states
near the $\Xi'_c \bar{K}$ threshold, which is close to the states $\Omega_c
(3065)$ and $\Omega_c (3050)$. For the isoscalar state with $1/2^-$, the
couplings of  channels $\Omega_c \eta$ and  $\Xi^*_c \bar{K}$ make the pole
deviate further from the threshold. The partial width to the $\Xi_c \bar{K}$ channel
is much smaller than $0.01$~MeV.  The mass of the isovector state $\Xi'_c
\bar{K}$ state with $1/2^-$ is strongly varied  by inclusion of the  $\Xi^*_c
\bar{K}$ or $\Omega_c \pi$ channel. And a considerably large width was mainly
provided by the coupling to the $\Omega_c\pi$ channel. Again, a very small
partial width to the $\Xi_c \bar{K}$ channel is found as in the isoscalar case.

{\noindent $\bullet$ \it Poles near the $\Xi_c \bar{K}$ threshold}

In the last part of Table~\ref{c45}, the poles near the $\Xi_c \bar{K}$
threshold are presented. For the isoscalar state with $1/2^-$, the position of
the pole is  motionless even after inclusion of all ten channels. Since the
$\Xi_c \bar{K}$ threshold is the lowest one among five channels, the width is
zero if no other lower channels are included. The same situation can be found in
the isoscalar case, though there exists two channels $\Omega_c^*\pi$ and
$\Omega_c\pi$ below the thresholds.

\renewcommand\tabcolsep{0.202cm}
\renewcommand{\arraystretch}{1}
\begin{table*}[hpbt!]    
\footnotesize
\caption{The masses and widths of molecular states from interactions in category
II near thresholds with a $\bar{K}^*$ at different values of $\alpha$ in the unit of GeV. Other notations are the same as Table~\ref{DDD}.
}\label{111}
\begin{tabular}{c|cr|rrrrrrrrrr}\toprule[1pt]
\multicolumn{13}{c}{Poles near the $\Xi^*_c \bar{K}^*$ threshold with $M_{th}=3540.2$~MeV}\\
\hline
$IJ^P$&$\alpha$  &$ cc$ &$ \Omega^*_c \omega$   &$\Xi_c^* \bar{K}^*$        &$\Omega_c \omega$   &$\Xi'_c \bar{K}^*$   &$\Xi_c \bar{K}^*$
&$ \Omega^*_c \eta$   &$\Omega_c \eta$                    &$\Xi_c^* \bar{K}$          &$\Xi'_c \bar{K}$          &$\Xi_c \bar{K}$ \\ \hline
                                        &$1.5$      &$0.0+0.1i$     &$1+0i$  &$0$&$0+0.0i$   &$0+0.0i$  &$0+0.0i$  &$0+0.0i$   &$0+0.0i$   &$0+0.0i$ &$0+0.0i$ &$0+0.0i$\\
   $0\frac{1}{2}^-$  &$2.0$     &$0.5+1.0i$    &$2+0i$  &$1$&$1+0.0i$  &$1+0.1i$  &$1+0.2i$   &$1+0.0i$   &$1+0.0i$   &$1+0.2i$  &$1+0.6i$  &$0.5+0.1i$   \\
                                  &$2.5$      &$0.7+2.3i$ &$6+0i$  &$2$ &$2+0.0i$  &$2+0.3i$   &$1+0.8i$  &$2+0.1i$    &$2+0.2i$  &$2+0.6i$  &$2+2.0i$  &$0.6+0.5i$      \\
                                     &$3.0$   &$0.0+3.4i$  &$10+0i$   &$4$ &$4+0.0i$  &$4+0.6i$   &$2+1.5i$  &$4+0.2i$    &$4+0.4i$  &$3+0.9i$  &$4+3.5i$  &$0.9+0.2i$  \\
\hline
                                          &$1.5$  &$1+0.4i$  &$1+0i$  &$0$&$1+0.0i$   &$1+0.1i$  &$0+0.0i$  &$1+0.4i$   &$1+0.0i$   &$0+0.0i$ &$0+0.0i$ &$0+0.0i$\\
   $0\frac{3}{2}^-$    &$2.0$    &$3+2.3i$ &$2+0i$ &$1$ &$2+0.0i$   &$2+0.2i$  &$0.5+0.4i$  &$3+2.2i$   &$2+0.4i$   &$1+0.0i$ &$1+0.2i$ &$1+0.0i$\\
                                         &$2.5$   &$5+4.8i$ &$5+0i$  &$3$&$4+0.1i$   &$4+0.7i$  &$1+1.4i$  &$7+4.4i$   &$4+1.0i$   &$3+0.1i$ &$3+0.4i$ &$3+0.2i$\\
                                          &$3.0$    &$7+7.7i$ &$8+0i$   &$5$ &$6+0.1i$   &$7+1.5i$  &$2+2.5i$  &$14+7.1i$   &$7+2.4i$   &$4+0.2i$ &$4+0.8i$ &$5+0.5i$\\
\hline
                          &$1.5$ &$1+1.4i$ &$1+0i  $  &$1$  &$1  +0.0i$     &$1  +0.1i$  &$1  +0.2i$  &$1 +0.6i$    &$1 +0.4i$   &$1 +0.0i$ &$1  +0.0i$ &$1  +0.0i$\\
   $0\frac{5}{2}^-$
                       &$2.0$ &$5+4.8i$ &$5+0i  $  &$5$  &$5  +0.0i$     &$5  +0.4i$  &$5  +0.6i$  &$6 +2.0i$   &$6 +1.4i$   &$5  +0.4i$ &$5  +0.4i$ &$5  +0.2i$\\
                 &$2.5$  &$15+11.2i$ &$11+0i$  &$11$ &$11+0.0i$   &$12+0.6i$  &$11+1.1i$ &$15+3.8i$  &$13 +3.6i$   &$9 +2.4i$ &$10+1.6i$ &$10 +0.4i$\\
                &$3.0$   &$25+21.0i$ &$19+0i$  &$18$ &$18+0.0i$   &$20+0.9i$  &$19+1.2i$ &$27+7.0i$  &$24+6.5i$   &$13+6.8i$   &$16+4.5i$ &$17+1.8i$\\
\hline
 $IJ^P$&$\alpha$ &$ cc$ &$ \Omega^*_c \rho$   &$\Xi_c^* \bar{K}^*$      &$\Xi'_c \bar{K}^*$&$\Omega_c \rho$  &$\Xi_c \bar{K}^*$
                                        &$\Xi_c^* \bar{K}$          &$\Xi'_c \bar{K}$          &$\Xi_c \bar{K}$        &$ \Omega^*_c \pi$    &$\Omega_c \pi$ \\ \hline


                                  &$2.5$    &$2+1.6i$         &$3+0i$  &$1$ &$1+0.0i$  &$1+0.0i$  &$1+0.0i$   &$1+0.0i$   &$1+0.2i$   &$1+0.0i$  &$1+0.0i$  &$1+0.0i$  \\
   $1\frac{1}{2}^-$ &$3.0$      &$6+3.4i$     &$8+0i$   &$2$ &$2+0.0i$  &$2+0.0i$   &$2+0.1i$  &$2+0.1i$    &$2+0.7i$  &$2+0.0i$  &$2+0.1i$  &$2+0.1i$  \\
                                    &$3.5$    &$11+5.2i$       &$10+0i$  &$4$&$5+0.1i$   &$5+0.1i$  &$5+0.3i$  &$3+0.3i$   &$4+1.7i$   &$4+0.0i$ &$3+0.3i$ &$5+0.2i$\\
\hline
                           &$2.5$ &$1+0.8i$ &$0.7+0i$  &$0.5$ &$0.5+0.0i$  &$0.5+0.0i$   &$0.5+0.0i$   &$0.5+0.0i$   &$0.5+0.0i$   &$0.5+0.0i$  &$0.5+0.2i$  &$0.5+0.1i $ \\
$1\frac{3}{2}^-$ &$3.0$ &$2+2.4i$&$1.7+0i$ &$0.7$ &$0.7+0.1i$  &$0.8+0.0i$  &$0.8+0.2i$&$0.6+0.1i$  &$0.7+0.0i$ &$0.7+0.0i$&$0.6+2.1i$  &$0.7+0.1i$\\
                             &$3.5$   &$5+4.5i$ &$3.5+0i$  &$1.1$&$1.1+0.2i$  &$1.6+0.1i$  &$1.4+0.5i$   &$0.7+0.5i$   &$1.1+0.0i$   &$1.1+0.0i$  &$1.2+4.2i$  &$1.2+0.4i$\\
                             &$4.0$   &$9+6.4i $ &$6.0+0i$  &$2.0$&$2.0+0.5i$  &$3.0+0.2i$  &$2.6+0.9i$   &$0.8+2.0i$   &$2.0+0.0i$   &$2.0+0.0i$  &$3.4+7.2i$  &$2.2+1.0i$\\
\hline

\multicolumn{13}{c}{Poles near the  $\Xi'_c \bar{K}^*$ threshold with $M_{th}=3470.7$~MeV}\\
\hline
$IJ^P$&$\alpha$       &$ cc$ &$ \Omega^*_c \omega$   &$\Xi_c^* \bar{K}^*$        &$\Omega_c \omega$   &$\Xi'_c \bar{K}^*$   &$\Xi_c \bar{K}^*$
&$ \Omega^*_c \eta$   &$\Omega_c \eta$                    &$\Xi_c^* \bar{K}$          &$\Xi'_c \bar{K}$          &$\Xi_c \bar{K}$ \\ \hline
                            &$1.5$    &$9+2.2i$&$3+0i$  &$6+0i$     &$5+0i$  &$3$    &$2+0.4i$    &$2  +0.0i$   &$3+1.0i$   &$2+0.2i$ &$3+0.1i$ &$3+0.1i$\\
   $0\frac{1}{2}^-$
                      &$2.0$   &$18+7.1i$  &$9+0i$  &$15+0i$  &$12+0i$&$8$    &$7 +1.3i$   &$7 +0.0i$   &$10+2.1i$   &$7+0.4i$  &$9+0.5i$  &$7+0.2i$   \\
                      &$2.5$&$24+17.8i$ &$14+0i$ &$24+0i$  &$20+0i$&$14$   &$11+2.4i$  &$13+0.0i$    &$17+3.0i$  &$13+1.5i$  &$15+1.2i$  &$11+0.7i$\\
                      &$3.0$  &$--$               &$20+0i$ &$37+0i$ &$34+0i$  &$19$ &$15+3.0i$  &$18+0.0i$    &$25+4.0i$  &$18+3.4i$  &$21+2.4i$  &$13+1.5i$\\
\hline
                         &$1.5$ &$8+2.4i$&$6 +0i$ &$8 +0i$   &$6 +0i$  &$6$  &$6  +0.3i$  &$6+0.0i$   &$ 6 +1.0i$   &$6 +0.1i$  &$4 +0.1i$ &$ 4 +0.1i$\\
   $0\frac{3}{2}^-$
                        &$2.0$ &$18+4.8i$&$14+0i$&$19+0i$ &$14+0i$&$14$  &$14+0.5i$ &$14+0.0i$ &$14+4.0i$ &$14+0.4i$ &$11+0.7i$ &$12+0.3i$\\
                      &$2.5$ &$30+10.3i$  &$24+0i$&$32+0i$   &$24+0i$&$24$  &$24+0.6i$  &$24+0.0i$   &$31+8.0i$ &$24+1.2i$ &$17+2.2i$ &$22+1.2i$\\
                   &$3.0$  &$43+15.7i$ &$35+0i$&$47+0i$   &$35+0i$&$35$  &$34+0.7i$  &$36+0.1i$   &$40+15.0i$ &$36+3.0i$ &$20+3.6i$ &$34+2.8i$\\
\hline
 $IJ^P$&$\alpha$ &$ cc$ &$ \Omega^*_c \rho$   &$\Xi_c^* \bar{K}^*$      &$\Xi'_c \bar{K}^*$&$\Omega_c \rho$       &$\Xi_c \bar{K}^*$
                                        &$\Xi_c^* \bar{K}$          &$\Xi'_c \bar{K}$          &$\Xi_c \bar{K}$        &$ \Omega^*_c \pi$    &$\Omega_c \pi$ \\ \hline
                                         &$2.0$&$6+2.1i$&$1 +0i$&$2 +0i$   &$0$  &$4  +0.0i$  &$0 +0.2i$  &$0 +0.0i$   &$0 +0.2i$   &$0 +0.2i$ &$0 +0.0i$ &$0 +0.6i$\\
  $1\frac{1}{2}^-$  &$2.5$ &$14+5.6i$ &$4 +0i$&$7 +0i$   &$4$ &$10+0.0i$ &$4+0.4i$  &$3  +0.1i$   &$3 +0.6i$   &$4 +0.4i$ &$3 +0.0i$ &$4 +1.5i$\\
                                      &$3.0$&$26+11.8i$ &$9+0i$&$14+0i$  &$8$ &$16+0.0i$ &$8+0.8i$  &$7 +0.3i$   &$7 +1.5i$   &$8 +0.7i$ &$7+0.1i$ & $9 +2.8i$\\
                                &$3.5$&$40+22.0i$&$15+0i$&$21+0i$ &$13$ &$26+0.0i$&$14+1.3i$ &$13+0.7i$ &$12+3.0i$ &$13+1.1i$ &$13+0.1i$ &$15+4.4i$\\
\hline
                                %
  $1\frac{3}{2}^-$    &$2.5$  &$3+3.5i $  &$0+0i$&$1+0i$ &$0$ &$1+0.0i$ &$0+0.1i$&$0+0.0i$   &$0+0.2i$   &$0+0.6i$&$0+0.0i$ &$0+1.5i$\\
                                          &$3.0$  &$6+5.9i$  &$1+0i$&$4+0i$   &$1$&$3+0.0i$  &$1+0.2i$  &$0+0.0i$   &$0+0.4i$   &$1+1.2i$ &$1+0.0i$ &$1+3.1i$\\
                                          &$3.5$  &$11+8.8i$ &$3+0i$&$7+0i$   &$3$  &$5+0.0i$&$3+0.3i$  &$2+0.1i$   &$2+0.7i$   &$2+2.0i$ &$3+0.0i$ &$3+5.8i$\\
                                          &$4.0$  &$17+13.0i $  &$4+0i$&$10+0i$   &$4$&$7+0.0i$  &$5+0.5i$  &$4+0.2i$   &$4+1.0i$   &$5+3.1i$ &$4+0.0i$ &$6+10.0i$\\
\hline
\multicolumn{13}{c}{Poles near the $\Xi_c \bar{K}^*$ threshold with $M_{th}=3363.3$~MeV}\\
\hline
$IJ^P$&$\alpha$     &$ cc$    &$ \Omega^*_c \omega$   &$\Xi_c^* \bar{K}^*$        &$\Omega_c \omega$   &$\Xi'_c \bar{K}^*$   &$\Xi_c \bar{K}^*$
&$ \Omega^*_c \eta$   &$\Omega_c \eta$                    &$\Xi_c^* \bar{K}$          &$\Xi'_c \bar{K}$          &$\Xi_c \bar{K}$ \\ \hline
                                &$1.5$ &$5+2.0i$ &$6 +0i$&$6 +0i$   &$6 +0i$ &$6  +0i$ &$6$ &$7  +0.0i$   &$7 +0.5i$   &$5  +2.2i$ &$5 +0.4i$ &$6  +0.0i$\\
   $0\frac{1}{2}^-$
                             &$2.0$ &$11+4.8i$&$14+0i$&$14+0i$ &$14+0i$ &$14+0i$ &$14$ &$15+0.0i$ &$17+0.7i$ &$12+6.4i$ &$11+1.3i$ &$14+0.0i$\\
                             &$2.5$&$16+10.5i$&$23+0i$&$23+0i$ &$23+0i$ &$23+0i$ &$22$ &$25+0.0i$ &$28+0.7i$ &$24+12.7i$ &$18+3.0i$ &$22+0.0i$\\
                            &$3.0$ &$23+27.6i$&$33+0i$&$34+0i$ &$33+0i$ &$35+0i$ &$32$ &$37+0.1i$   &$43+0.6i$   &$40+20.6i$ &$23+5.8i$ &$32+0.0i$\\
\hline
                               &$1.5$ &$4+2.4i$&$6 +0i$&$6 +0i$   &$6 +0i$ &$6  +0i$ &$6$ &$7  +0.1i$   &$6 +0.0i$   &$5 +2.0i$ &$6 +0.4i$ &$6  +0.0i$\\
   $0\frac{3}{2}^-$
                        &$2.0$&$9+4.8i$&$14+0i$&$14+0i$ &$14+0i$ &$14+0i$ &$14$ &$16+0.2i$ &$14+0.0i$ &$13+6.4i$ &$13+1.5i$ &$14+0.0i$\\
                        &$2.5$ &$17+7.8i$ &$23+0i$&$23+0i$ &$23+0i$ &$23+0i$ &$22$ &$27+0.4i$   &$23+0.0i$   &$23+8.1i$ &$23+3.3i$ &$22+0.0i$\\
                       &$3.0$&$29+14.3i$&$34+0i$&$34+0i$ &$33+0i$ &$35+0i$ &$32$ &$40+0.6i$   &$32+0.2i$   &$36+12.8i$ &$34+6.0i$ &$32+0.0i$\\
\hline
 $IJ^P$&$\alpha$ &$ cc$ &$ \Omega^*_c \rho$   &$\Xi_c^* \bar{K}^*$      &$\Xi'_c \bar{K}^*$&$\Omega_c \rho$       &$\Xi_c \bar{K}^*$
                                        &$\Xi_c^* \bar{K}$          &$\Xi'_c \bar{K}$          &$\Xi_c \bar{K}$        &$ \Omega^*_c \pi$    &$\Omega_c \pi$ \\ \hline
                                        &$2.0$  &$1+2.7i$   &$1 +0i$&$1 +0i$   &$1  +0i$  &$1 +0i$&$1$  &$1 +0.2i$   &$1 +0.1i$   &$1 +0.0i$ &$1 +0.7i$ &$1 +0.6i$   \\
  $1\frac{1}{2}^-$
                                        &$2.5$   &$5+7.0i$  &$3 +0i$&$2 +0i$   &$2+0i$ &$2+0i$ &$2$ &$3  +0.5i$   &$2 +0.4i$   &$2 +0.0i$ &$4 +1.6i$ &$3 +1.9i$  \\
                                          &$3.0$ &$11+13.0i$&$6+0i$&$5+0i$  &$5+0i$ &$5+0i$&$4$  &$7 +1.0i$   &$4 +0.6i$   &$4 +0.0i$ &$8+2.3i$ & $7 +3.8i$\\
                                          &$3.5$&$21+21.7i$ &$10+0i$&$8 +0i$ &$8 +0i$&$8 +0i$ &$7$&$11 +1.7i$ &$7 +0.9i$ &$7 +0.0i$ &$13+2.9i$ &$13+4.5i$\\
\hline
                           &$2.0$ &$2+4.0i$            &$1 +0i$   &$1 +0i$   &$1  +0i$  &$1 +0i$&$1$  &$1 +0.2i$   &$1 +0.2i$   &$1 +0.0i$ &$1 +0.8i$ &$1 +0.2i$   \\
  $1\frac{3}{2}^-$
                           &$2.5$  &$5+9.0i$          &$3 +0i$    &$2 +0i$   &$2+0i$ &$2+0i$ &$2$ &$3  +0.5i$   &$3 +0.5i$&$2 +0.0i$ &$4 +1.9i$ &$2 +0.4i$  \\
                           &$3.0$&$6+17.0i$          &$6+0i$    &$5+0i$  &$5+0i$ &$5+0i$&$4$  &$7 +1.0i$   &$6 +0.9i$&$4 +0.0i$ &$8+3.0i$ & $5 +0.5i$\\
                            &$3.5$&$5+18.0i$        &$10+0i$  &$8+0i$ &$8+0i$&$8+0i$ &$7$&$11+1.4i$ &$9+1.5i$ &$7+0.0i$ &$14+4.3i$ &$8+0.6i$\\
\hline
\bottomrule[1pt]
\end{tabular}
\end{table*}

{\noindent $\bullet$ \it Poles near the $\Xi^*_c \bar{K}^*$ threshold}

In Table~\ref{111}, the poles near the $\Xi^*_c \bar{K}^*$ threshold are
presented.In the calculation, ten channels are involved, which provide nonzero
width. Besides acquiring a width of several MeV, the isoscalar state with spin
parity $1/2^-$ becomes very shallow, and even disappears when the $\alpha$
increases to about $3.0$ due to  the repulsive coupled-channel effect in
ten-channel calculation. The strongest couplings can be found in the $\Xi'_c
\bar{K}$ channel, and considerable coupled-channel effects are also found in
channels $\Xi_c\bar{K}^*$ and $\Xi_c^*\bar{K}$. The isoscalar state with $3/2^-$
becomes a little tight after inclusion of the coupled-channel effect, and a
width of about 15~MeV can be acquired with a mass gap  about 7~MeV to the
threshold at an $\alpha$ value of 3. The two-channel calculations suggest that
the largest coupled-channel effect is from the $\Omega^*_c\eta$ channel, and
channels $\Xi'_c\bar{K}^*$, $\Xi_c\bar{K}^*$, and $\Omega_c\eta$ also provide
considerable contributions. For the isoscalar state with $5/2^-$, a large width
about 40~MeV can be reached at an $\alpha$ value of about 3. No obvious dominant
channel can be found in the two-channel calculations, and four channels
$\Omega^*_c\eta$, $\Omega_c\eta$, $\Xi_c^*\bar{K}$, and $\Xi'_c\bar{K}$ show their
strong couplings to this state.

In the isovector sector, there exist two poles with spin parities $1/2^-$ and
$3/2^-$. For the state with $1/2^-$, the most dominant channel is found in the
$\Xi'_c\bar{K}$ channel. However, the variation of the mass is mainly from the
inclusion of the $\Omega^*_c\rho$ channel.  For the state with $3/2^-$, the
$\Omega_c^*\pi$ channel provides a large width while the variation of mass is also
mainly from the $\Omega^*_c\rho$ as the case with $1/2^-$.

{\noindent $\bullet$ \it Poles near  the $\Xi'_c \bar{K}^*$ threshold}

In the second part of Table~\ref{111}, the results for four poles near the
$\Xi'_c \bar{K}^*$ threshold are listed.  In the isoscalar case, the states with
spin parity $1/2^-$ from the single-channel calculation deviate further from
the threshold and acquire a nonzero width.  However, with the increase of the
$\alpha$ value, the pole becomes dim in the complex plane. The couplings of the
state to the channels $\Omega^*_c\omega$ and $\Omega^*_c\eta$ are very weak. All
other channels have considerable impact on the mass or width of
the state. For the state with $3/2^-$, the $\Xi^*_c\bar{K}$ channel plays an
important role in the variation of the mass, while  nonzero width is mainly provided
 by the $\Omega_c\eta$ channel.

For the isovector state with $1/2^-$, the total width is much larger than the
sum of the partial widths from the two-channel calculations. It should be from
the couplings between the channels besides these involved in the two-channel
calculation. For the state with spin parity $3/2^-$, the variation of the mass
is mainly from the $\Xi^*_c\bar{K}^*$ channel, and the $\Omega_c\pi$ channel provides
the most important contribution to the total width.

{\noindent $\bullet$ \it Poles near the $\Xi_c \bar{K}^*$ threshold}

The results for the pole near the $\Xi_c \bar{K}^*$ threshold are present as the
last part of Table~\ref{111}. In the single-channel calculation, the $\Xi_c
\bar{K}^*$ states with different $J^P$ are almost degenerate. After including
the coupled-channel effect, the degeneration disappears for all four states. For
example, the mass gap between two isoscalar states reaches 5~MeV at an $\alpha$
value of 3. The differences of the widths of isoscalar and isovector states are
also obvious.

In the isoscalar sector, the state with $1/2^-$ becomes more shallow after
including the coupled-channel effect, which is mainly from the couplings to the
$\Xi'_c\bar{K}$ channel. A width about 10~MeV can be reached at $\alpha$ value
of 2, which is mainly from the contribution of the $\Xi^*_c\bar{K}$ channel. For
the state with $3/2^-$, the $\Xi^*_c \bar{K}$ channel is dominant to
produce its total width. And the $\Xi'_c\bar{K}$ channel also couples to the state strongly.

In the isovector case,  the mass of state with $1/2^-$ decreases obviously after
including all channels. However, none of two-channel calculations exhibits such
behavior. It should be from the couplings of the channels except the $\Xi_c
\bar{K}^*$   channel. For the same reason, the total width is also much
larger than the sum of widths from two-channel calculations. The  couplings to
the channels  $\Omega^*_c\pi$ and $\Omega_c\pi$ are found to be important.  The state
with $3/2^-$ is shallow even at an $\alpha$ value of 3.5. Its total width is
smaller than the sum of the partial widths as in the case with $1/2^-$.

\section{Summary and discussion\label{sec5}}

The molecular states generated from the meson-baryon channels with quantum
numbers $C=1$, $ S=-2$ under $3.6$~GeV are studied in a quasipotential
Bethe-Salpeter equation approach.  With the help of the light meson exchange
model, the single-channel and coupled-channel calculations are performed. Based
on the results obtained in the current work, the following conclusions can be
reached:
\begin{enumerate}[leftmargin=*,itemsep= 0 pt]

  \item[$\bullet$] $\Omega_c(3120)$: Both isoscalar and isovector states can be
  produced from the  $\Xi_c^* \bar{K}$ interaction  with spin parity $3/2^-$.
  The latter requires larger $\alpha$ value than the former, which can be
  produced at an $\alpha$ value of about 1. The state with $3/2^-$ has very weak
  coupling to the $\Xi_c\bar{K}$ channel, which is the observation channel of
  the $\Omega_c(3120)$. Besides, its width increases very rapidly to about
  10~MeV with the increase of the parameter. Hence, the isoscalar $\Xi_c^*
  \bar{K}$ state is a better candidate of $\Omega_c(3120)$. The theoretical width
  seems small compared with the experimental value, which may be due to
  absence of  light decay channels. Besides,  the small partial width of
  $\Xi_c\bar{K}$ channel may be the reason why it is difficult to observe in
  the experiment considering that a molecular state itself should be more hardly
  produced than a conventional $\Omega_c$ state. Generally speaking, the
  isoscalar $\Xi_c^* \bar{K}$ state withe $3/2^-$ is a good assignment of the
  $\Omega_c(3120)$. Higher-precision measurement should be helpful to confirm
  it. Besides, an isovector state is also suggested to be observed in the
  $\Omega_c\pi$ channel based on the current calculation.

  \item[$\bullet$] $\Omega_c(3050)$ and $\Omega_c(3065)$: These two states are
  close to the $\Xi'_c\bar{K}$ threshold. Based on the calculations in the
  current work and
  literature~\cite{Karliner:2017kfm,Wang:2017zjw,Padmanath:2017lng,Debastiani:2017ewu},
  molecular states can be produced from $\Xi'_c\bar{K}$ interaction with
  $1/2^-$. However,  current calculation suggests extremely weak couplings of
  these states to the experimental observation channel $\Xi_c\bar{K}$. Combined
  with the rejection of the spin parity $1/2^-$ suggested in
  Ref.~\cite{LHCb:2021ptx}, the observed $\Omega_c(3050)$ and $\Omega_c(3065)$
  should not be the molecular states from the $\Xi'_c\bar{K}$ interaction.
  However, it is suggested to search for an isovector state in the $\Omega_c\pi$
  channel.

  \item[$\bullet$] The molecular states with higher mass were predicted in the
  current works. We suggest the experimental research of such states in the
  future experiment, especially the followings states. Near the $\Xi^*D^*$ threshold, 
  it is suggested to search the states with $I(J^P)=0(1/2^-)$ and $0(5/2^-)$ in the $\Xi^*D$ channel,
  and the state with $0(3/2^-)$ in the $\Xi D^*$ channel. The state near the $\Xi^*D$ threshold  with $0(3/2^-)$ couples strongly to the  $\Xi D^*$ channel.  Near the $\Xi^*_c\bar{K}$ threshold, the states  with $0(1/2^-)$  and $1(3/2^-)$ are suggested to be searched  in the $\Xi'_c\bar{K}$ channel,  and the state with $0(3/2^-)$ in the
  $\Omega^*_c\eta$ channel.  The state near the $\Xi'_c\bar{K}$ threshold with
  $0(1/2^-)$ couples strongly to the $\Omega_c\eta$ channel while that with $1(3/2^-)$ decays mainly to the $\Xi\bar{K}$  channel.  The states near the $\Xi_c\bar{K}^*$ threshold with $0(1/2^-)$ and
  $0(3/2^-)$ can be searched in the $\Xi^*_c\bar{K}$ channel.  \end{enumerate}


\vskip 10pt

\noindent {\bf Acknowledgement} This project is supported by the National Natural Science
Foundation of China (Grants No. 11675228).

\end{document}